%% file: main.tex
\documentclass[11pt,journal]{IEEEtran}
\usepackage{etoolbox}

\usepackage{multirow}
\usepackage{siunitx}
\usepackage{booktabs} 
\usepackage{array}    
\usepackage{colortbl} 
\usepackage[table,xcdraw]{xcolor} 
\definecolor{lightgray}{gray}{0.95} 
\definecolor{headergray}{gray}{0.8} 
\usepackage{placeins}
\usepackage{url}
\usepackage{textcomp}
\usepackage[most]{tcolorbox}
\usepackage{diagbox}
\usepackage{fixltx2e}
\usepackage[table]{xcolor}
\usepackage{cite}
\usepackage{float}
\usepackage{subcaption}
\usepackage{textcomp}
\usepackage{graphicx}
\usepackage[skip=2pt,font=small]{caption}
\usepackage{mathabx}
\usepackage{algorithmic}
\usepackage{array}
\usepackage{mdwmath}
\usepackage{mdwtab}
\usepackage{eqparbox}
\usepackage{hyperref}

\usepackage{bm}
\usepackage{longtable}
\usepackage{lscape} 

\usepackage[first=0,last=9]{lcg}

\begin{document}
\graphicspath{ {images/} }

\title{Comprehensive Evaluation of OCT-based Automated Segmentation of Retinal Layer, Fluid and Hyper-Reflective Foci: Impact on Clinical Assessment of Diabetic Retinopathy Severity\\}

\author{
    Shuo Chen\IEEEauthorrefmark{1},
    Da Ma\IEEEauthorrefmark{2},
    Munispriyan Raviselvan\IEEEauthorrefmark{3},
    Sathishkumar Sundaramoorthy\IEEEauthorrefmark{3},\\
    Karteek Popuri\IEEEauthorrefmark{8},
    Myeong Jin Ju\IEEEauthorrefmark{4}\IEEEauthorrefmark{5},
    Marinko V. Sarunic\IEEEauthorrefmark{6}\IEEEauthorrefmark{7},\\
    Dhanashree Ratra\IEEEauthorrefmark{3},
    and Mirza Faisal Beg\IEEEauthorrefmark{1}%
    
    \thanks{\IEEEauthorrefmark{1}School of Engineering Science, Simon Fraser University, Burnaby, BC, Canada.}%
    \thanks{\IEEEauthorrefmark{2}School of Medicine, Wake Forest University, Winston-Salem, NC, United States.}%
    \thanks{\IEEEauthorrefmark{3}Sankara Nethralaya, Chennai, Tamil Nadu, India.}%
    \thanks{\IEEEauthorrefmark{4}Department of Ophthalmology and Visual Sciences, The University of British Columbia, Vancouver, BC, Canada.}%
    \thanks{\IEEEauthorrefmark{5}School of Biomedical Engineering, The University of British Columbia, Vancouver, BC, Canada.}%
    \thanks{\IEEEauthorrefmark{6}Institute of Ophthalmology, University College London, London, United Kingdom.}%
    \thanks{\IEEEauthorrefmark{7}Department of Medical Physics and Biomedical Engineering, University College London, London, United Kingdom.}%
    \thanks{\IEEEauthorrefmark{8}Department of Computer Science, Memorial University of Newfoundland, St. John's, Canada.}%
    \thanks{Corresponding authors: Mirza Faisal Beg (faisal-lab@sfu.ca), Da Ma (dma@wakehealth.edu), Shuo Chen (shuo\_chen\_4@sfu.ca).}
}

\maketitle

\begin{abstract}

\textit{Background}: Diabetic retinopathy (DR) is a major cause of vision loss, and early detection is essential to prevent irreversible blindness. Spectral Domain Optical Coherence Tomography (SD-OCT) enables high-resolution retinal imaging, while AI-driven segmentation improves diagnostic precision. However, segmentation performance varies across models, especially for DR cases with differing severity and complex fluid and hyperreflective foci (HRF) patterns. The clinical deployment of these models remains underexplored. This study develops an active-learning-based deep learning pipeline for the automated segmentation of retinal layers, fluid, and HRF, comparing state-of-the-art (SOTA) models and evaluating their impact on DR assessment.

\textit{Methods}: Four deep learning models (U-Net, SegFormer, SwinUNETR, VM-UNet) were trained on manually annotated SD-OCT volumes to segment ten retinal layers, fluid, and HRF. Five-fold cross-validation assessed segmentation performance. Retinal thickness was quantified using a K-nearest neighbours (KNN) algorithm and visualized via Early Treatment Diabetic Retinopathy Study (ETDRS) maps. Structural differences between Non-Proliferative (NPDR) and Proliferative DR (PDR) were analyzed, including correlations with visual acuity. 

\textit{Results}: SwinUNETR achieved the highest overall accuracy (DSC = 0.7719; NSD = 0.8149), while VM-UNet outperformed in specific layers. PDR showed increased RNFL thickness and fluid accumulation, whereas NPDR exhibited thickening in GCL+IPL and OS. In NPDR, thickening in RNFL, INL, OPL, EZ, and the accumulation of fluid and HRF correlated with reduced vision. In PDR, RNFL thickening, GCL+IPL and ONL+IS thinning and fluid accumulation were associated with visual impairment.

\textit{Conclusion}: 
The proposed pipeline enables accurate, efficient DR analysis with reduced manual effort. SwinUNETR and VM-UNet performed robustly in complex regions, though HRF segmentation remains challenging. Thickness maps generated from auto-segmentation offer clinically relevant insights, supporting improved disease monitoring and treatment planning.

\end{abstract}


\section{Introduction}\label{Intro}
Diabetic retinopathy (DR) is a prevalent microvascular complication of diabetes mellitus (DM) and a leading cause of vision impairment worldwide \cite{KobrinKlein2007OverviewRetinopathy}. It was estimated that approximately 20\% of diabetic individuals over the age of 50 will develop DR, which, if left untreated, can progress to severe visual impairment or blindness \cite{Teo2021Global2045}. The progression of DR was influenced by several risk factors, including prolonged diabetes duration, elevated glycated hemoglobin (HbA1c) levels, hypertension, hyperlipidemia, obesity, and smoking \cite{Zhou2025CoffeeStudy, Mori2025AssociationDiabetes, Cheung2010DiabeticRetinopathy, Zhang2024AssociationPatients, Cai2018TheMeta-analysis}. Clinically, DR is categorized into two primary stages: Non-Proliferative Diabetic Retinopathy (NPDR) and Proliferative Diabetic Retinopathy (PDR). NPDR represents an early stage, often asymptomatic, characterized by microvascular abnormalities that progressively compromise retinal capillary integrity. Without timely medical intervention, NPDR can advance to PDR, a more severe stage marked by pathological neovascularization due to chronic retinal ischemia. This progression increases the risk of severe complications such as vitreous hemorrhage and retinal detachment, ultimately threatening vision.

Spectral Domain Optical Coherence Tomography (SD-OCT) is a cutting-edge, non-invasive imaging technique that provides high-resolution, cross-sectional visualization of retinal structures. Its real-time image acquisition capability makes it an invaluable tool for DR screening and early diagnosis \cite{GabrieleE.Lang2007OpticalRetinopathy}. SD-OCT enables the detection of subclinical retinal changes by quantifying variations in retinal thickness and identifying fluid accumulation. Accurate and reliable segmentation of retinal layers and pathological features is crucial for DR diagnosis and treatment planning. 

Numerous studies have explored automated segmentation techniques for retinal layer analysis. Herzog \textit{et al.} proposed an edge maximization and smoothness-constrained thresholding approach to delineate retinal boundaries \cite{Herzog2004RobustTomography}. Chiu \textit{et al.} utilized a graphical cut algorithm to minimize the weighted sum of edge paths along connected nodes, effectively segmenting retinal layers \cite{Chiu2010AutomaticSegmentation}. Wang \textit{et al.} introduced a multi-step approach that includes artifact removal, contrast enhancement, and segmentation via level set methods, k-means clustering, and Markov random fields (MRFs) \cite{Wang2015SegmentationImages}. Traditional machine-learning techniques have also been employed for fluid segmentation. González \textit{et al.} identified dark fluid regions in OCT scans using support vector machines (SVM) and random forest classifiers \cite{Gonzalez2013AutomaticAnalysis}. Chen \textit{et al.} applied a graph-cut classifier followed by a region-growing algorithm for cystoid macular edema (CME) segmentation \cite{XinjianChen2012Three-DimensionalGraph-Search-Graph-Cut}. However, these conventional approaches are limited by their reliance on handcrafted features and their susceptibility to performance degradation in severely diseased cases.

Deep learning has emerged as a powerful alternative for automated retinal segmentation, offering greater robustness against variations in image quality and pathological abnormalities. Liu \textit{et al.} utilized a ResNet-based convolutional neural network (CNN) combined with a random forest classifier for patch-wise layer segmentation \cite{Liu2019AutomatedClassifier}. Kugelman \textit{et al.} proposed a recurrent neural network (RNN) with a graph search framework to segment retinal layers in both healthy individuals and patients with age-related macular degeneration (AMD) \cite{Kugelman2018AutomaticSearch}. Hu \textit{et al.} developed a multi-scale CNN capable of capturing different feature levels for improved segmentation accuracy \cite{Hu2019AutomaticSearch}. U-Net and its derivatives have become widely adopted among deep-learning models for medical image segmentation. U-Net's encoder-decoder architecture, enhanced by skip connections, enables efficient spatial information preservation and mitigates vanishing gradient issues \cite{Ronneberger2015U-Net:Segmentation}. It has been successfully applied to retinal layer segmentation, fluid detection, and HRF analysis, achieving state-of-the-art (SOTA) performance \cite{Roy2017ReLayNet:Networks,Ma2021LF-UNetImages,Tennakoon2018RetinalNetworks,Schlegl2018FullyImages}. Generative adversarial networks (GANs) were also used for retinal boundary augmentation and segmentation adaptation cross multiple OCT domains\cite{Kugelman2023EnhancedCross-localisation, Chen2023Segmentation-guidedNetworks}. Vision Transformers (ViTs) have recently outperformed CNNs in large-scale datasets. Unlike CNNs, which rely on local receptive fields, ViTs employ self-attention mechanisms to capture global dependencies, which is particularly beneficial for detecting diffuse fluid regions. Xue \textit{et al.} implemented a Swin-Transformer-based architecture for fluid segmentation in diabetic macular edema (DME) and AMD, demonstrating superior performance over traditional CNN-based models \cite{Xue2024RetinalSwin-Unet}. Kulyabin \textit{et al.} leveraged the Segment Anything Model (SAM) for retinal fluid segmentation, incorporating point and bounding box prompts to outperform U-Net in macular hole and fluid segmentation tasks \cite{Kulyabin2024SegmentBiomarkers}. Despite these advancements, most existing studies focus on either the retinal layer or fluid segmentation, with varying levels of segmentation performance on pathological clinical features. However, limited efforts are dedicated to investigating the effect of automated segmentation performance on NPDR/PDR classification or prognosis, which is crucial to evaluating their clinical translation.

Studies have examined the relationship between retinal layer thickness, fluid accumulation, and DR severity. Browning \textit{et al.} analyzed macular thickness across different DR severity levels and observed a correlation between macular thickening and increased risk of subclinical edema \cite{Browning2008TheEdema}. Kim \textit{et al.} investigated choroidal thickness alterations in DR and DME patients, reporting a significant increase in choroidal thickness as DR severity progressed from mild/moderate NPDR to PDR \cite{Kim2013ChangesPatients}. Cho \textit{et al.} assessed macular and peripapillary retinal thickness in DR subjects, identifying statistically significant differences in retinal thickness across seven anatomical regions between DR and control groups \cite{Cho2010DiabeticThickness}. Santos \textit{et al.} demonstrated that fluid accumulation within the outer segment (OS) layer is significantly associated with central retinal thickness and visual impairment in DME patients \cite{SantosT2024VisionLayer}. These findings suggest that retinal layer thickness and fluid distribution are both reliable biomarkers for DR diagnosis and progression monitoring. However, limited efforts are dedicated to investigating the effect of automated segmentation performance on DR/PDR classification or prognosis, which is crucial to evaluating their clinical translation.

The current study introduces an end-to-end framework integrating retinal layer and fluid segmentation with a statistical analysis of structural changes in DR patients. The key contributions include:
\begin{enumerate}
    \item Development of an efficient active-learning-based segmentation pipeline for severely pathological DR patients.
    \item Comprehensive evaluation of multiple SOTA deep learning models, revealing differential performance on segmenting retinal layers, fluid, and HRF segmentations, using both volume- and thickness-based evaluation metrics for all biomarkers, differentially considering under-segmentation and over-segmentation cases.
    \item Evaluate the clinical translatability of the auto-segmentation-based retinal thicknesses, fluid and HRF biomarkers for differentiating DR severity, as well as their association with visual acuity.
\end{enumerate}

\section{Methods}\label{Methods}

\subsection{Data Acquisition}\label{dataacq}
116 SD-OCT volumes were acquired from Sankara Nethralaya Eye Care Hospital in India. The imaging data were obtained using the Cirrus HD-OCT 5000 (Carl Zeiss Meditec, Dublin, CA, USA). Seventeen OCT volumes were captured in Macular Cube mode with a 512 × 128 pixel resolution, while the remaining 99 volumes were scanned in OCTA mode at 350 × 350 pixels. Both modes covered a 6 × 6 mm² macular region centered on the fovea. Despite differences in scanning speed and resolution, Wong \textit{et al.} reported no significant variation in macular thickness measurements between the two modes \cite{Wong2024Comparison5000}. Table \ref{tab:demographic_info} presents the demographic details of the subjects in two DR severity groups, showing no significant differences in age, diabetes duration, or visual acuity (p $>$ 0.05). However, the gender distribution differs due to the limited number of female patients in the PDR group. The variance inflation factor (VIF) is calculated for DR groups, age, gender, duration of diabetes, and visual acuity. No significant multicollinearity is found as all values are close to 1. 

\begin{table*}[htb]
\centering
\setlength{\tabcolsep}{10pt} 
\renewcommand{\arraystretch}{1.3} 
\begin{tabular}{@{}lcccccc@{}}
\rowcolor{headergray} 
\textbf{Group} & \textbf{N} & \textbf{Age (Mean ± SD)} & \textbf{Gender} & \textbf{Duration of Diabetes (yrs)} & \textbf{Visual Acuity (LogMAR)} \\ 
\midrule
\rowcolor{lightgray} 
NPDR & 66 & 60.00 ± 8.85 & Female: 34, Male: 32 & 15.94 ± 7.97 & 0.35 ± 0.35 \\ 
PDR & 50 & 56.84 ± 8.24 & Female: 13, Male: 37 & 14.66 ± 8.60 & 0.47 ± 0.41 \\ 
\midrule
\rowcolor{lightgray} 
p-value & - & 0.05023 & 0.00457 & 0.41508 & 0.10351 \\ 
VIF & - & 1.158384 & 1.840780 & 1.162841 & 1.642617 \\
\bottomrule
\end{tabular}
\caption{Demographic information of experimental DR groups. Numerical values are presented as mean ± standard deviation (SD). The visual acuity is expressed in the logarithm of the Minimum Angle of Resolution (LogMAR). The p-values are calculated using the Welch's t-test. The variance inflation factor (VIF) is calculated for each variable, including the DR category.}
\label{tab:demographic_info}
\end{table*}

\subsection{Pre-processing}\label{dataprep}
To prepare the raw OCT volumes for further analysis, we performed several pre-processing steps:
\begin{itemize}
\item The approximate retinal center in each B-scan was adjusted to align with the center along the axial direction. The axial retinal center was estimated by computing the average axial position of pixels whose axial intensity values are more significant than the lowest 20th percentile. This helped initialize a starting point for axial motion correction. 
\item A 3D Bounded Variation (BV) smoothing technique was applied to suppress noise while preserving smoother structural boundaries, providing better contrasts for manual labelling and model prediction. 
\item Motion artifacts among adjacent B-scans were corrected in both the axial and lateral directions. Axial translations were determined through cross-registration using the moving average of the central B-scan as a reference. Lateral corrections were achieved by performing registration based on the adjacent B-scans' discrete Fourier transform (DFT). Rotational adjustments were computed by transforming translations into polar space, using the moving average of the central B-scan as a reference.
\end{itemize}

\subsection{Active-Learning-Based Ground Truth Segmentation Annotation}\label{dataseg}
\begin{figure}[htb]
\centering
\includegraphics[width=\linewidth]{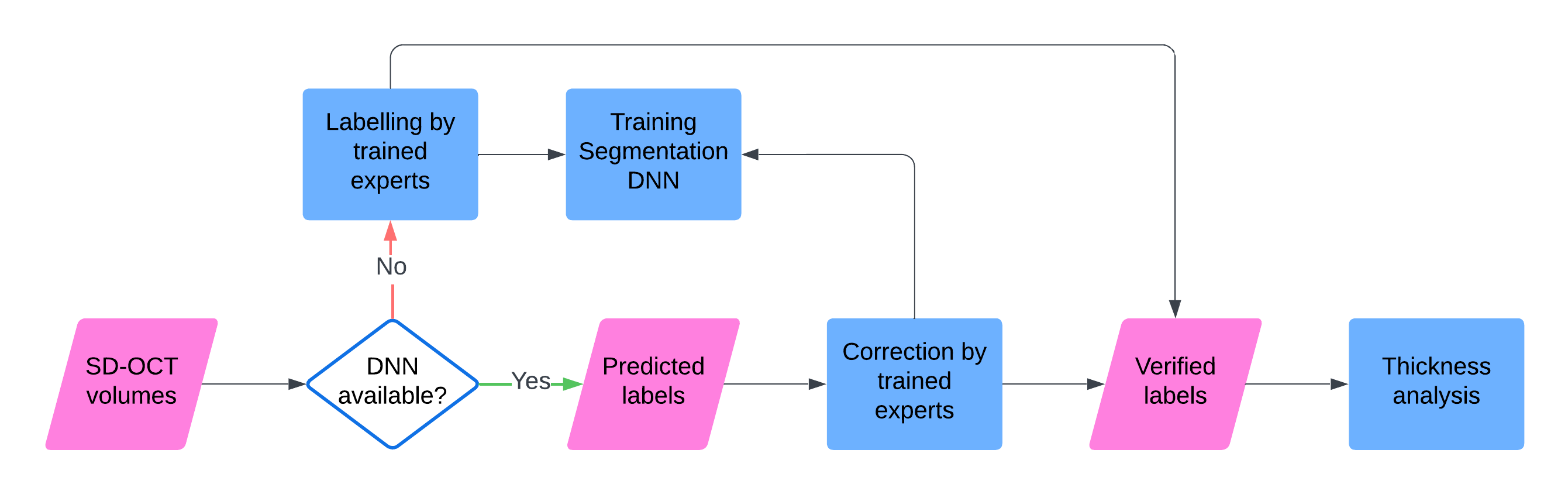}
\caption{Manual segmentation pipeline. Multiple iterations were performed between DNN training and manual corrections. A thickness analysis was conducted after segmentation has been completed and verified.}
\label{fig:flowchart}
\end{figure}

Figure \ref{fig:flowchart} illustrates that the active-learning-based semi-automatic segmentation follows a structured human-in-the-loop (HITL) interactive labelling workflow. SwinUNETR was used solely to generate initial auto-segmentations, which were subsequently corrected by human annotators, thereby improving labelling efficiency. However, all models, including SwinUNETR, were trained and evaluated from scratch using the finalized, expert-corrected labels. As such, the initial use of SwinUNETR does not introduce any bias or discrepancies in the quantitative evaluation of model performance.
Initially, five volumes were manually annotated from scratch. Manual segmentation was performed on every fifth B-scan, while the intermediate B-scans were interpolated under the assumption that adjacent B-scans share structural similarities. However, B-scans that exhibited significant structural changes were individually labelled and corrected. These five manually labelled volumes served as the first iteration of the training dataset to train a deep neural network (DNN) with a data split ratio of 3:1:1 for training, validation, and testing. The initially trained model was then used to generate segmentation predictions for an additional 20 volumes, which were subsequently reviewed and manually corrected. This iterative process continued with a 7:2:1 data split ratio in subsequent iterations for training, validation, and testing, with the network being retrained on an expanded dataset each time. This ensured that all volumes underwent accurate segmentation and manual verification. Volume splits were stratified to ensure pathological cases with all label types (i.e. retinal layer, fluid, and HRF) were presented in training, validation, and testing sets.

\subsection{Automatic Segmentation Networks Architecture}\label{datanet}
We investigated the performance of four deep neural network (DNN) architectures, which are either widely used for medical image segmentation or have demonstrated SOTA performance in related tasks:
\begin{itemize}
\item \textbf{U-Net}: The most widely-used, well-established medical image segmentation model that employs a CNN-based encoder-decoder architecture with skip connections \cite{Ronneberger2015U-Net:Segmentation}. While effective, it may struggle with high-resolution inputs due to a lack of global contextual awareness. The U-Net model is configured with a depth of five channels and three residual units.
\item \textbf{SegFormer}: A transformer-based architecture proposed by Xie \textit{et al.}, which utilizes a hierarchical transformer encoder combined with a lightweight MLP decoder to enhance feature extraction \cite{Xie2021SegFormer:Transformers}. The 2D variant of SwinUNETR is used for training B-scans.
\item \textbf{SwinUNETR}: A CNN-Transformer-composited architecture proposed by Hatamizadeh \textit{et al.}, which replaces the CNN-based encoder in U-Net with a Swin Transformer encoder, enabling multi-scale feature extraction through a shifted windowing mechanism. \cite{Hatamizadeh2022SwinImages}. This is also the architecture that is used for the semi-automatic generation of the ground truth segmentation labels through the HITL active-learning process. 
\item \textbf{VM-UNet}: A recently proposed novel architecture developed by \textit{Ruan et al.}, this model introduces a state-space model (SSM) and an asymmetric encoder-decoder structure. It models the visual data as an evolving state, efficiently capturing both local and global dynamics with a structure inspired by continuous dynamic systems, balancing computational efficiency while maintaining a global contextual view \cite{Ruan2024VM-UNet:Segmentation}.
\end{itemize} 

\subsection{Segmentation Model Training}\label{datatrain}
We employed 5-fold cross-validation, stratified by DR diagnosis, with a 4:1:1 ratio for training, validation, and testing. Each input consisted of a 3-channel image constructed by three repetitions of a single B-scan. Each training B-scan was resized to $512 \times 512$. To mitigate class imbalance among segmentation labels, the excess Vitreous and Choroid regions were cropped. Various augmentation techniques were applied, including lateral flipping, Gaussian noise injection($\mu = 0,\, \sigma^2 = 0.01$), random contrast enhancement ($\gamma \in (0.5, 4.5)$), rotation within the B-scan plane (± 20°), random zooming ($(0.5, 1.5)$) and random intensity shifting (± 10).

For loss functions, we used combinations of Dice loss, cross-entropy (CE) loss, and L1 loss for texture differences. Given the ground truth label $y$ and predicted label $\hat{y}$, for every pixel $i$, the Dice loss is calculated as:

\begin{equation}
L_{dice}(y, \hat{y}) = 1 - \frac{2 \cdot \sum_{i} y_i \hat{y_i}}{\sum_{i} y_i^2 + \sum_{i} \hat{y_i}^2 + \epsilon}
\end{equation}

We set $\epsilon$ to $10^{-6}$ to avoid the division by zero problem. The CE loss is defined as:

\begin{equation}
\resizebox{0.9\columnwidth}{!}{$
L_{CE}(y, \hat{y}) = -\frac{1}{N} \sum_{i} \left[ y_i \log(\hat{y}_i) + (1 - y_i) \log(1 - \hat{y}_i) \right]
$}
\end{equation}

The Sobel operator calculates gradients in horizontal($G_x$) and vertical($G_y$) directions, and the total gradient magnitude $G$ is the Euclidean norm. Given the label $Y$, the gradients are calculated as:

\begin{equation*}
\resizebox{0.9\columnwidth}{!}{$
G_x(Y) = Y \ast \begin{bmatrix}
-1 & 0 & 1 \\
-2 & 0 & 2 \\
-1 & 0 & 1
\end{bmatrix}, \quad
G_y(Y) = Y \ast \begin{bmatrix}
-1 & -2 & -1 \\
0 & 0 & 0 \\
1 & 2 & 1
\end{bmatrix}
$}
\end{equation*}

\begin{equation}
G(Y) = \sqrt{G_x(Y)^2 + G_y(Y)^2}
\end{equation}

The texture loss is defined as the L1 norm between the predicted and ground truth labels:

\begin{equation}
L_{texure}(y, \hat{y}) = \frac{1}{N} \sum_{i} \left| G_i(y) - G_i(\hat{y}) \right|
\end{equation}

Thus, the total loss function is calculated as:

\begin{equation}
L(y, \hat{y}) = \alpha \cdot L_{dice} + \beta \cdot L_{CE} + \gamma \cdot L_{texure}
\end{equation}

The $\alpha$, $\beta$, and $\gamma$ are weighting factors for Dice, CE and texture losses, respectively. We empirically set $\alpha = \beta = \gamma = 1$ for our experiment. 

We empirically assigned different class weights to CE loss to emphasize the class imbalance issue. Specifically, we assigned 0.1 to Vitreous and Choroid, 0.5 to the rest of the layers, and 1 to fluid and HRF. We used AdamW optimizer with CosineAnnealing scheduler with the warm restart. We adopted the distributed parallel learning supported by the PyTorch Lightning module\footnote{https://lightning.ai/}, with a batch size of 8 and a learning rate of 1e-4. The training was deployed on an NVIDIA V100 Volta GPU allocated by Cedar Compute Canada\footnote{More information can be found at: https://docs.alliancecan.ca/wiki/Cedar}. 

\subsection{Segmentation Performance Evaluation}\label{dataeval}
We evaluated the segmentation performance by overlapping areas and boundary alignment. We used the Dice similarity coefficient (DSC) to measure the similarity between the predicted and ground truth masks. Given correctly predicted pixels as True Positives(TP), incorrectly predicted pixels as False Positives(FP), and missing predicted pixels as False Negatives(FN), the Dice score is calculated as:

\begin{equation}
Dice = \frac{2 \cdot TP}{2 \cdot TP + FP + FN}
\end{equation}

Nikolov \textit{et al.} proposed the normalized surface Dice (NSD) to estimate the deviation of surface contours within a certain threshold $\tau$\cite{Nikolov2018DeepRadiotherapy}. Defining a set of Euclidean distances from predicted segmentation $\hat{Y}$ to ground truth segmentation $Y$ as $\mathcal{D}_{\hat{Y}Y}$, and vice versa as $\mathcal{D}_{Y\hat{Y}}$, we obtain the subset of distances that are smaller or equal to the threshold $\tau$ as:

\begin{equation*}
\mathcal{D}_{\hat{Y}Y}^\prime = \{d \in \mathcal{D}_{\hat{Y}Y} | d \leq \tau\}
\end{equation*}

\begin{equation}
\mathcal{D}_{Y\hat{Y}}^\prime = \{d \in \mathcal{D}_{Y\hat{Y}} | d \leq \tau\}
\end{equation}

The NSD is calculated as :

\begin{equation}
NSD = \frac{|\mathcal{D}_{\hat{Y}Y}^\prime| + |\mathcal{D}_{Y\hat{Y}}^\prime|}{|\mathcal{D}_{\hat{Y}Y}| + |\mathcal{D}_{Y\hat{Y}}|}
\end{equation}

Special attention is needed for fluid evaluation. For True Negative(TN) cases where both ground truth and predicted fluid are absent, the NSD score should be the correct prediction. For False Positive(FP) and False Negative(FN) cases where the fluid is only present in one of the ground truths or predicted segmentations, the NSD score should be zero as an incorrect prediction. We set $\tau$ to $10$ pixels for all classes, roughly $3\%$ of the shortest image edge. The model performance will be evaluated without any of the post-processing steps mentioned in the original papers. 

Additionally, we defined the Under-Segmentation Score (USS) and the Over-Segmentation Score (OSS) to evaluate if the model fails to detect certain regions or assigns excessive labels to a class. Given the confusion matrix for N classes:
\begin{equation}
CM = \begin{bmatrix}
    \text{TP}_1 & \text{FP}_{1,2} & \text{FP}_{1,3} & \dots & \text{FP}_{1,N} \\
    \text{FN}_{2,1} & \text{TP}_2 & \text{FP}_{2,3} & \dots & \text{FP}_{2,N} \\
    \text{FN}_{3,1} & \text{FN}_{3,2} & \text{TP}_3 & \dots & \text{FP}_{3,N} \\
    \vdots & \vdots & \vdots & \ddots & \vdots \\
    \text{FN}_{N,1} & \text{FN}_{N,2} & \text{FN}_{N,3} & \dots & \text{TP}_N
\end{bmatrix}
\end{equation}

We computed the USS and OSS for certain class C as: 
\begin{equation}
    USS_C = \frac{\sum_{\forall j \neq C} \text{CM}[C, j]}{\sum \text{CM}[C, :]}
\end{equation}

\begin{equation}
    OSS_C = \frac{\sum_{\forall i \neq C} \text{CM}[i, C]}{\sum \text{CM}[:, C]}
\end{equation}

A higher USS score indicates that a significant portion of the ground truth class $C$ is not detected, leading to under-segmentation, while a higher OSS score suggests that the model over-predicts class $C$, leading to over-segmentation. We used a heuristic cutoff value of 0.2 to determine if there is under-segmentation or over-segmentation.

\subsection{Retinal Layer Thickness Analysis}\label{datathickness}
Layer thickness computation was performed using the K-Nearest Neighbors (K-NN) algorithm. The layer boundaries were converted into 3D point clouds. For each data point on the upper layer, the closest corresponding point on the lower layer was identified based on Euclidean distance. The distance is properly adjusted by the voxel dimension along each axis. The thickness maps are resized to the resolution of $350 \times 350$ for consistent representation. The vitreous and choroid layers were excluded from these calculations due to their unbounded nature on one side. Given that the inner retinal layers—including RNFL, GCL+IPL, INL, OPL, and ONL+IS—converge at the foveal pit, the central region was excluded from their thickness analysis to ensure more reliable and anatomically consistent measurements.

\begin{figure}[htb]
\centering
\includegraphics[width=0.6\columnwidth, keepaspectratio]{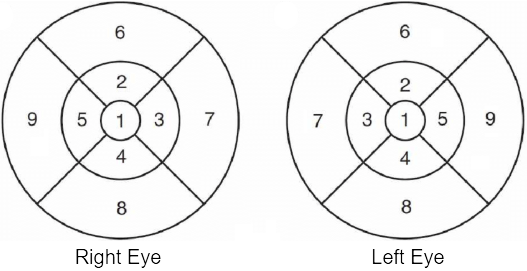}
\caption{ETDRS diagram for both left and right eyes. The diameters of the central circle, inner ring, and outer ring are 1 mm, 3 mm, and 6 mm, respectively. Nine subfields are numbered and named as follows: 1-C(Central field), 2-SI(Superior Inner), 3-NI(Nasal Inner), 4-II(Inferior Inner), 5-TI(Temporal Inner), 6-SO(Superior Outer), 7-NO(Nasal Outer), 8-IO(Inferior Outer), 9-TO(Temporal Outer). }
\label{fig:etdrs}
\end{figure}

The Early Treatment Diabetic Retinopathy Study (ETDRS) grid was employed to assess thickness variations systematically across different macular regions. As depicted in Figure \ref{fig:etdrs}, this grid divides the macula into three concentric circles with diameters of 1\si{mm}, 3\si{mm}, and 6\si{mm}, all centered on the fovea. These circles define the central, inner, and outer subfields, subdivided into four quadrants: superior, inferior, nasal, and temporal.

\subsection{Statistical Analysis}\label{statistic}
The DSC and NSD scores were calculated for model segmentation performance, and the thickness measurements were derived from the predicted segmentation. The mean DSC and NSD scores were compared within each retinal region. We used a generalized linear model (GLM) to assess the segmentation performance, the thickness difference between the NPDR and PDR groups, and the correlation between the thickness and visual acuity within each DR group. 

Specifically, we compared the DSC and NSD scores of each pair of models with Gaussian distributions, and the performance is ranked via the effect size and p-values after the false discovery rate (FDR) correction. We compared the thickness differences between NPDR and PDR groups in each layer sector while adjusting for relevant covariates, including age, gender, and duration of diabetes. Visual acuity was not included as it is considered a downstream clinical outcome rather than a demographic or biological confounder. The compound Poisson-Gamma distribution was used to model zero-inflated and highly skewed thickness measurements across DR groups while controlling for age, gender, and duration of diabetes. The correlation between the visual acuity and layer sector thickness within each DR group was modelled using the Gaussian distribution while controlling for age, gender, and duration of diabetes. We converted each categorical variable to numerical values. We assigned 0 to NPDR and 1 to the PDR group, and assigned 0 to females and 1 to males. The models' estimated coefficients (beta values) along with their 95\% confidence intervals (CI) were calculated and visualized. Statistically significant results before and after FDR correction were explicitly highlighted.

\section{Results}\label{results}
\subsection{Segmentation}\label{resultseg}
\begin{figure}[htb]
\centering
\includegraphics[width=\linewidth]{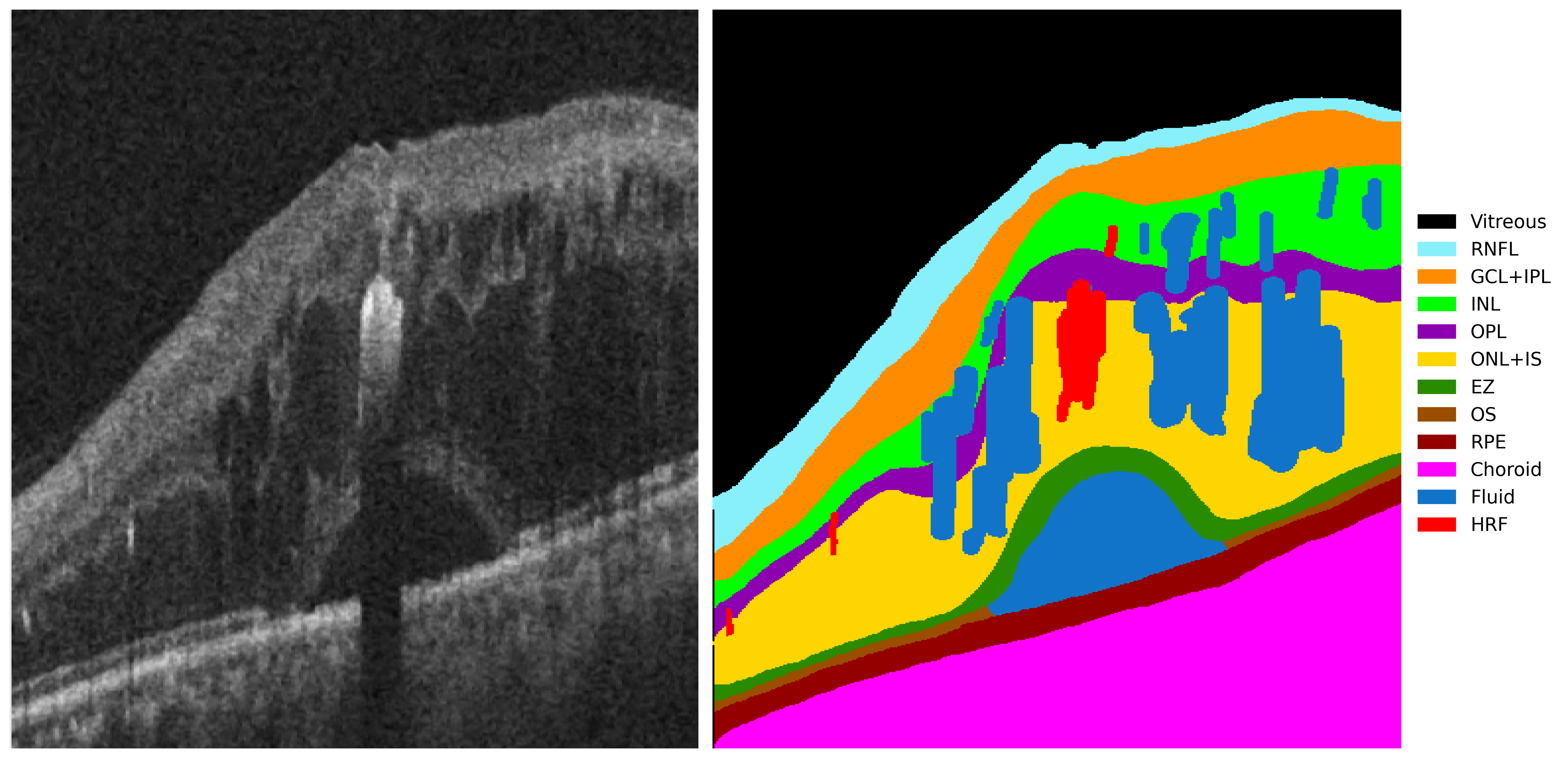}
\caption{Example of retinal OCT and ground truth segmentation derived through the active-learning pipeline (image on the right) for a single B-scan(image on the left). This image was acquired from a 59-year-old female patient with NPDR. Ten retinal layers were segmented from top to bottom, plus the fluid and HRF within the retinal body.}
\label{fig:oct_seg}
\end{figure}

Figure \ref{fig:oct_seg} illustrates a representative SD-OCT B-scan with ground truth retinal layer and fluid segmentation derived from the active-learning-based HITL semi-automatic segmentation pipeline. The segmentation delineates nine essential retinal layers: the Retinal Nerve Fiber Layer (RNFL), Ganglion Cell Layer and Inner Plexiform Layer (GCL+IPL), Inner Nuclear Layer (INL), Outer Plexiform Layer (OPL), Outer Nuclear Layer and Inner Segment Layer (ONL+IS), Ellipsoid Zone (EZ), Outer Segment Layer (OS), and Retinal Pigment Epithelium (RPE). The region above the Internal Limiting Membrane (ILM) is also identified as the Vitreous, while the Choroid lies beneath Bruch's Membrane (BM). Fluid segmentation involves three primary fluid types: intraretinal fluid (IRF), subretinal fluid (SRF), and pigment epithelial detachment (PED), all of which appear as hypo-reflective regions between the ILM and BM. Furthermore, hyperreflective foci (HRF), which manifest as high-intensity dot-like or clustered lesions, are also segmented.

Table \ref{tab:dice_seg} presents the segmentation results for four models, with values averaged across five-fold cross-validation. Tables \ref{tab:dice} and \ref{tab:nsd} separately report the DSC and NSD metrics to quantify segmentation volume overlap and boundary distance, respectively. SwinUNETR achieves the highest overall DSC and NSD among the evaluated models, demonstrating superior segmentation performance, particularly in the OPL, Choroid, and HRF regions. VM-UNet exhibits competitive performance, achieving the best DSC and NSD scores in the Vitreous, RNFL, and fluid regions. U-Net and SegFormer perform comparably, though U-Net slightly outperforms SegFormer in DSC across most layers, whereas SegFormer demonstrates marginally better NSD performance. These findings suggest that model predictions are less consistent in these layers, potentially due to structural complexity or segmentation challenges inherent to these regions.

\begin{table*}[htb]
    \centering
    \small 
    \renewcommand{\arraystretch}{1.2} 
    \setlength{\tabcolsep}{6pt} 

    \begin{subtable}[b]{\textwidth}
        \centering
        \resizebox{0.95\textwidth}{!}{ 
        \begin{tabular}{@{}lcccccccccc|cc|c@{}}
        \toprule
        \diagbox{Model}{Dice}{Label} & Vitreous & RNFL & GCL+IPL & INL & OPL & ONL+IS & EZ & OS & RPE & Choroid & Fluid & HRF & Avg. \\
        \midrule
        U-Net & 0.9887 & 0.8723 & 0.8928 & 0.8180 & 0.7714 & 0.9151 & \textbf{0.7247} & \textbf{0.7331} & \textbf{0.8510} & 0.9800 & 0.2522 & 0.4075 & 0.7672 \\
        SegFormer & 0.9897 & 0.8722 & 0.8932 & 0.8157 & 0.7680 & 0.9139 & 0.7210 & 0.7254 & 0.8403 & 0.9772 & 0.2122 & 0.3228 & 0.7543 \\
        SwinUNETR & 0.9871 & 0.8713 & 0.8961 & 0.8259 & \textbf{0.7788} & 0.9148 & 0.7175 & 0.7267 & 0.8440 & \textbf{0.9804} & 0.2806 & \textbf{0.4402} & \textbf{0.7719} \\
        VM-UNet & \textbf{0.9899} & \textbf{0.8740} & \textbf{0.8988} & \textbf{0.8269} & 0.7768 & \textbf{0.9185} & 0.7120 & 0.7173 & 0.8396 & 0.9798 & \textbf{0.2813} & 0.4211 & 0.7697 \\
        \bottomrule
        \end{tabular}
        }
    \caption{Mean Dice Similarity Coefficient (DSC)}
    \label{tab:dice}
    \end{subtable}

    \vspace{1.2em} 

    \begin{subtable}[b]{\textwidth}
        \centering
        \resizebox{0.95\textwidth}{!}{ 
        \begin{tabular}{@{}lcccccccccc|cc|c@{}}
        \toprule
        \diagbox{Model}{NSD}{Label} & Vitreous & RNFL & GCL+IPL & INL & OPL & ONL+IS & EZ & OS & RPE & Choroid & Fluid & HRF & Avg. \\
        \midrule
        U-Net & 0.9540 & 0.8860 & 0.8408 & 0.8555 & 0.8381 & 0.8414 & 0.9338 & \textbf{0.9252} & \textbf{0.8786} & 0.9160 & 0.2847 & 0.5187 & 0.8061 \\
        SegFormer & 0.9604 & 0.8888 & 0.8455 & 0.8580 & 0.8438 & 0.8434 & \textbf{0.9343} & 0.9231 & 0.8700 & 0.9181 & 0.2253 & 0.4105 & 0.7934 \\
        SwinUNETR & 0.9576 & 0.8890 & \textbf{0.8499} & \textbf{0.8661} & \textbf{0.8493} & \textbf{0.8497} & 0.9313 & 0.9199 & 0.8726 & \textbf{0.9207} & 0.3186 & \textbf{0.5546} & \textbf{0.8149} \\
        VM-UNet & \textbf{0.9628} & \textbf{0.8899} & 0.8484 & 0.8644 & 0.8425 & 0.8452 & 0.9327 & 0.9190 & 0.8586 & 0.9167 & \textbf{0.3221} & 0.5418 & 0.8120 \\
        \bottomrule
        \end{tabular}
        }
    \caption{Mean Normalized Surface Dice (NSD)}
    \label{tab:nsd}
    \end{subtable}

    \caption{Comparison of segmentation performance across four models. Dice and NSD scores were calculated by averaging over five validation folds. The "Average" column represents the mean performance across all retinal regions per model. The best score in each layer is highlighted in \textbf{bold}.}
    \label{tab:dice_seg}
\end{table*}

\begin{figure*}[htb]
\centering
\includegraphics[width=0.85\textwidth, height=0.85\textheight, keepaspectratio]{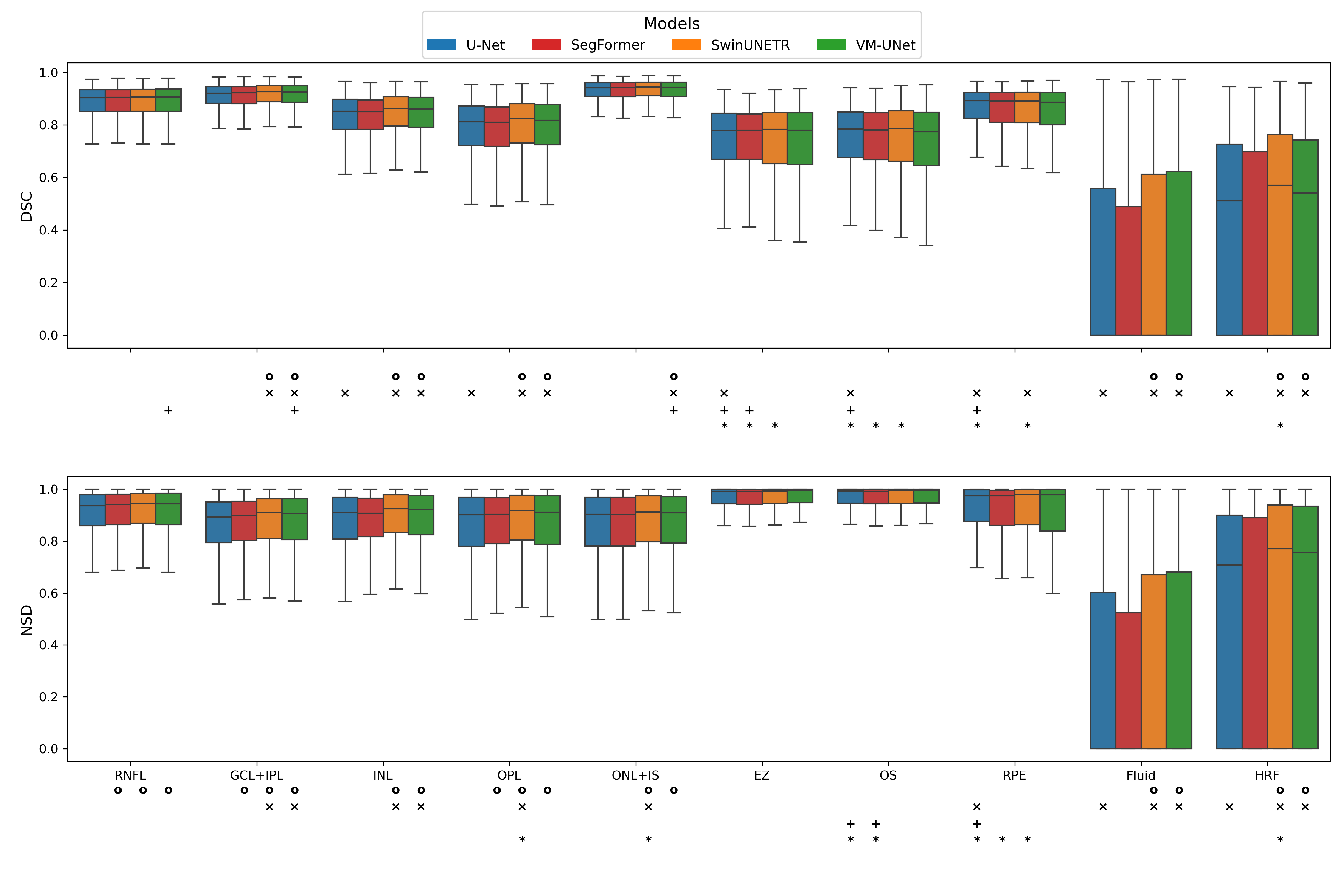}
\caption{Comparison of segmentation performance across models using Area Dice (top) and Surface Dice (bottom) metrics for each retinal region (Vitreous and Choroid are excluded). Each boxplot summarizes the Dice scores for U-Net, SegFormer, SwinUNETR, and VM-UNet across all subjects. Statistical significance between models is assessed using GLM with FDR correction. Significance markers below each group indicate which models significantly outperform others: 'o' indicates significantly better than U-Net, '×' better than SegFormer, '+' better than SwinUNETR, and '*' better than VM-UNet.}
\label{fig:dice_glm}
\end{figure*}

Figure \ref{fig:dice_glm} presents a comparative analysis of segmentation performance across U-Net, SegFormer, SwinUNETR, and VM-UNet using DSC (top row) and NSD (bottom row) across various retinal regions, excluding Vitreous and Choroid. Each boxplot illustrates the distribution of Dice metric per model, while the statistical significance of pairwise differences is determined using GLM with FDR correction. Significance markers below each region indicate statistically superior performance relative to other models: circles (‘o’) for U-Net, crosses (‘×’) for SegFormer, plus signs (‘+’) for SwinUNETR, and asterisks (‘*’) for VM-UNet. SwinUNETR and VM-UNet demonstrate consistent improvements over baseline models in several regions, notably in DSC and NSD for GCL+IPL, INL, Fluid, and HRF. However, U-Net and SegFormer show significantly better performance in the DSC of the EZ layer and NSD of the OS layer compared to SwinUNETR and VM-UNet. Additionally, SwinUNETR significantly outperforms VM-UNet in DSC for EZ, OS, RPE, and HRF, as well as NSD for OPL, ONL+IS, RPE, and HRF. Conversely, VM-UNet shows significantly better performance than SwinUNETR in DSC for RNFL and GCL+IPL. Quantitative results are shown in Supplementary Tables \ref{tab:seg_glm_DSC} and \ref{tab:seg_glm_NSD}.

\begin{figure*}[htbp]
    \centering
    \captionsetup{justification=centering}

    \begin{subfigure}[b]{0.45\textwidth}
        \centering
        \includegraphics[width=\linewidth, keepaspectratio]{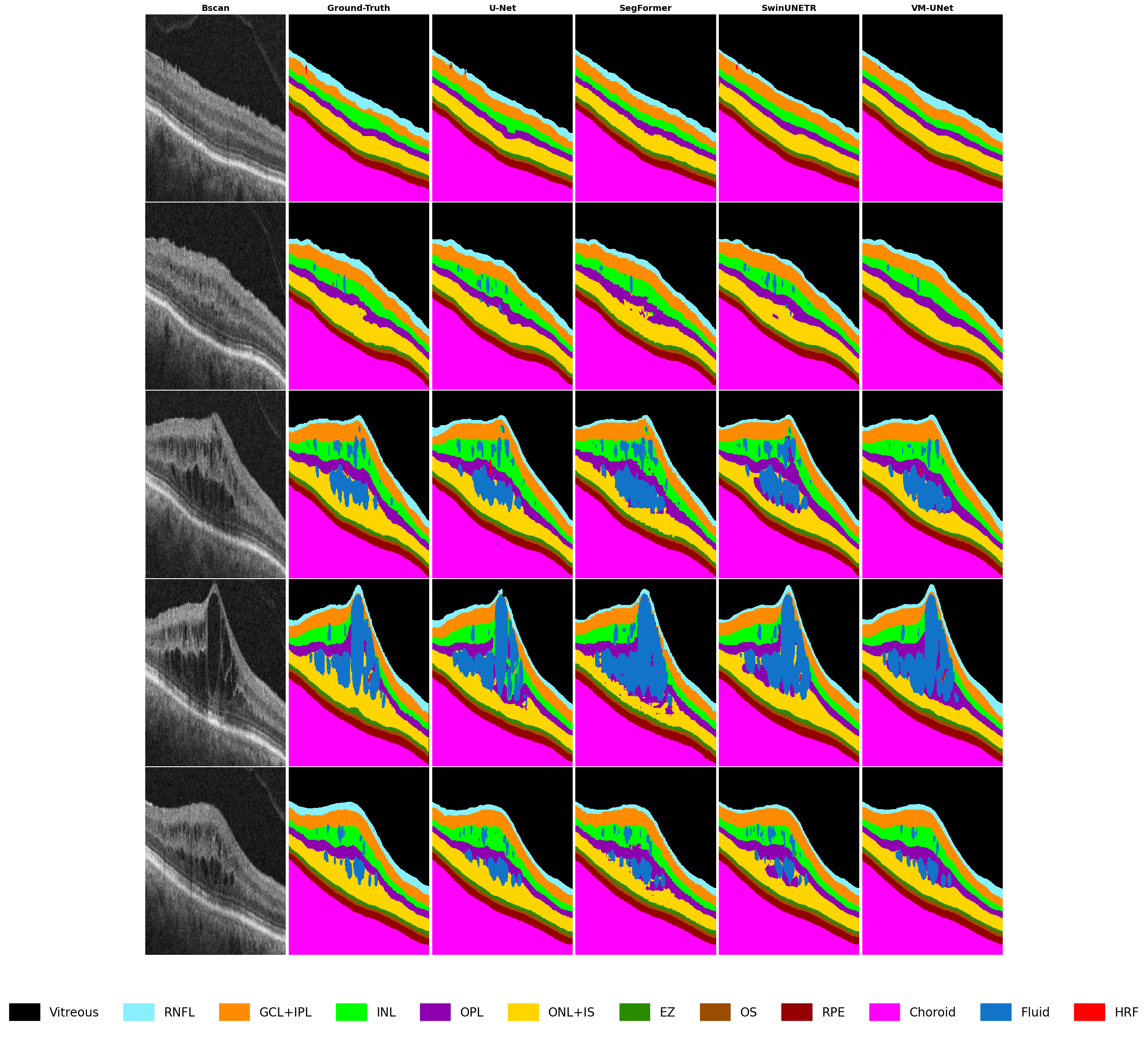}
        \caption{NPDR patient with severe fluid accumulation}
        \label{fig:seg_pics1}
    \end{subfigure}
    \begin{subfigure}[b]{0.45\textwidth}
        \centering
        \includegraphics[width=\linewidth, keepaspectratio]{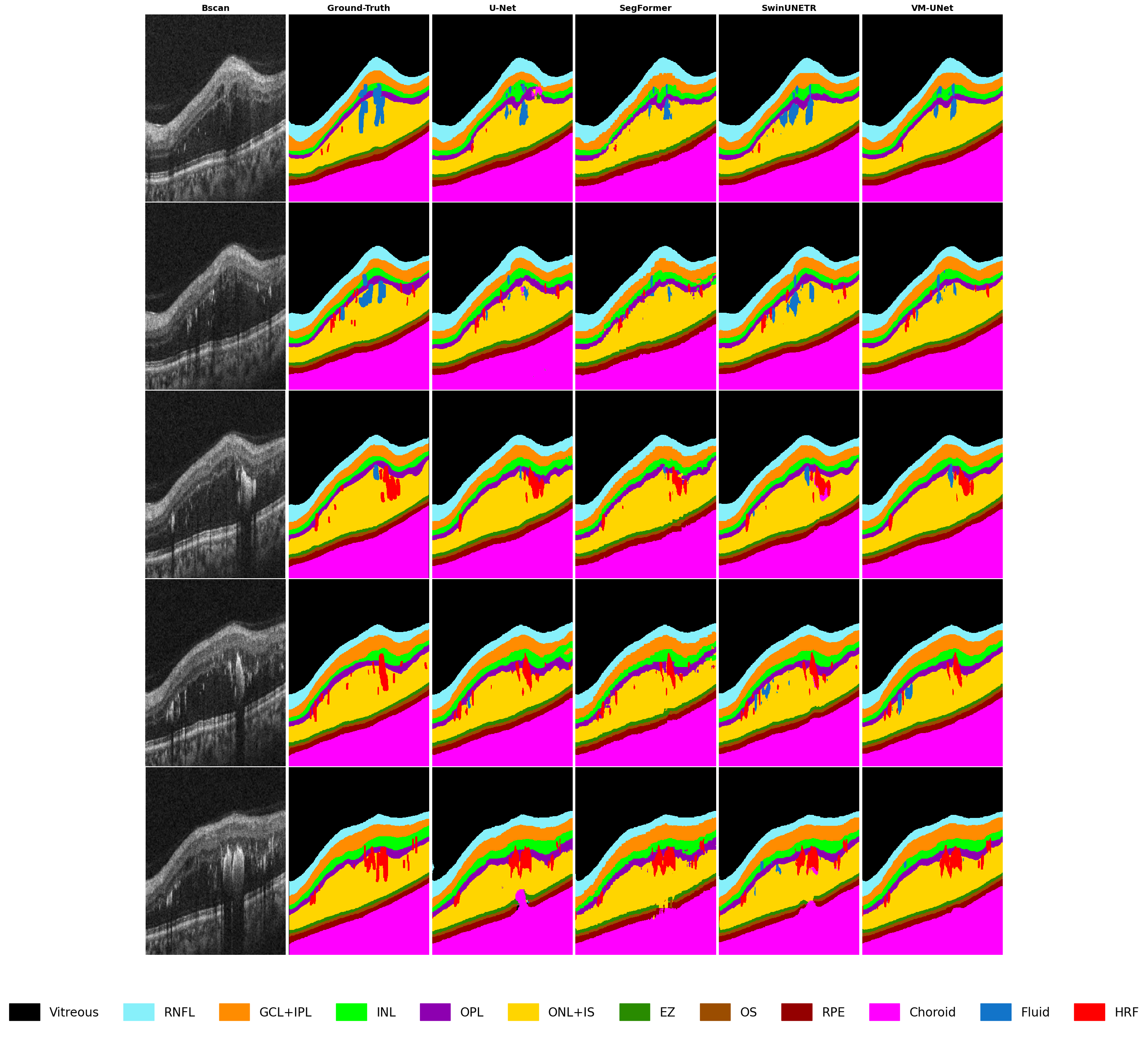}
        \caption{NPDR patient with HRF and shading artifacts.}
        \label{fig:seg_pics2}
    \end{subfigure}

    \begin{subfigure}[b]{0.45\textwidth}
        \centering
        \includegraphics[width=\linewidth, keepaspectratio]{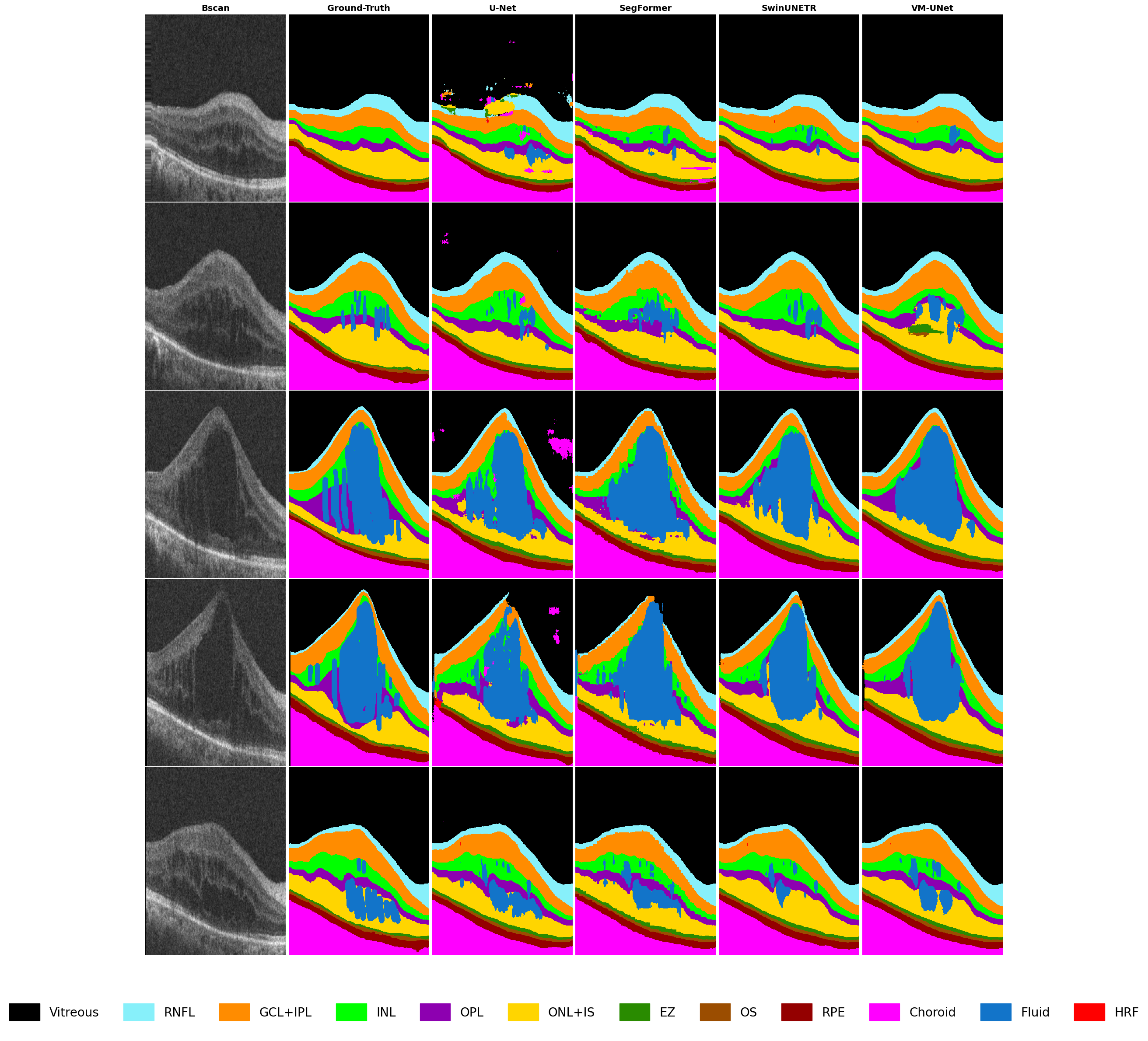}
        \caption{PDR patient with severe fluid accumulation.}
        \label{fig:seg_pics3}
    \end{subfigure}
    \begin{subfigure}[b]{0.45\textwidth}
        \centering
        \includegraphics[width=\linewidth, keepaspectratio]{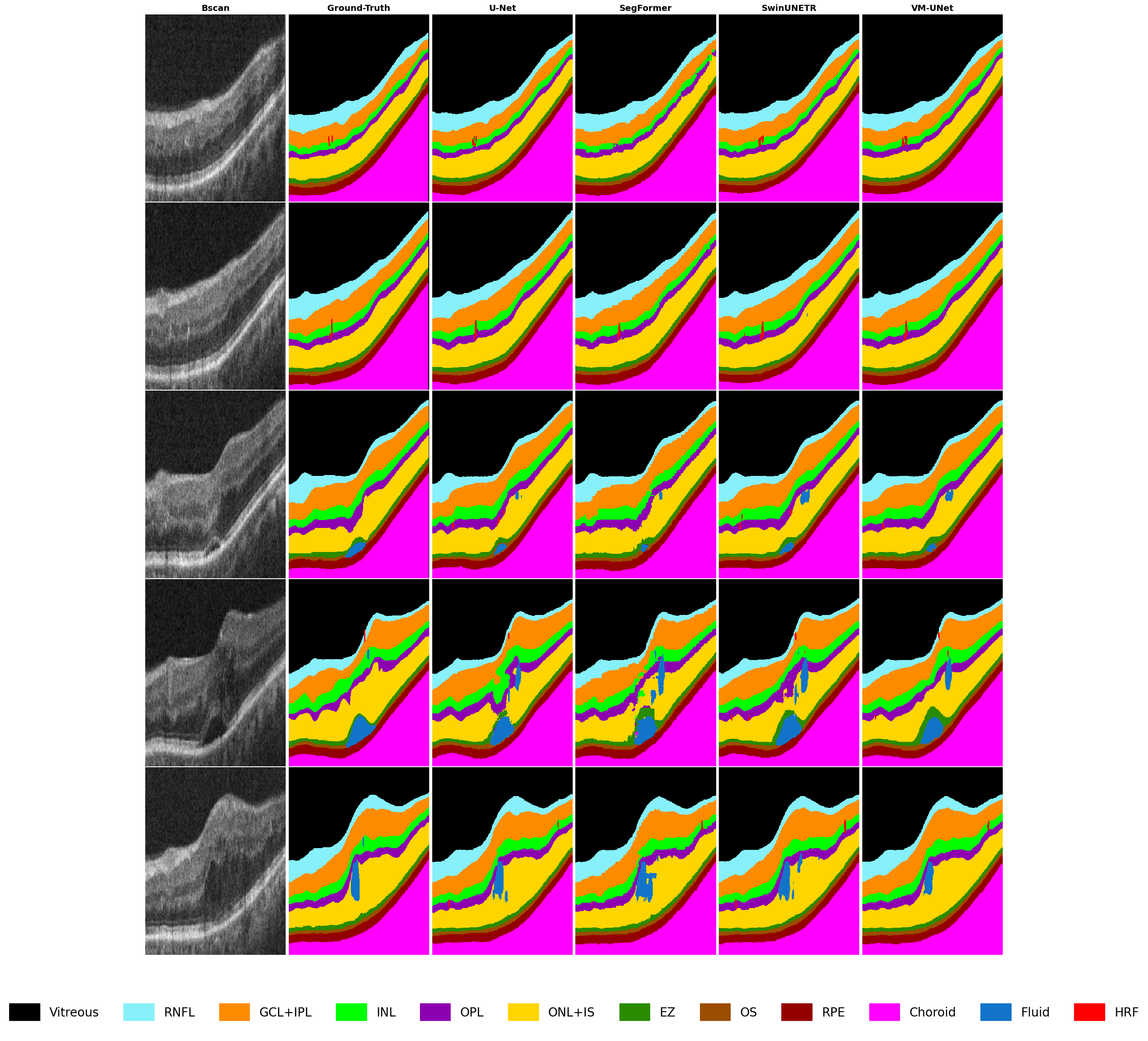}
        \caption{NPDR patient with subretinal fluid (SRF).}
        \label{fig:seg_pics4}
    \end{subfigure}

    \caption{Comparison of OCT B-scan segmentation results across different retinal conditions. Each row represents a different B-scan, while columns correspond to different segmentation models and patient conditions.}
    \label{fig:seg_pics}

\end{figure*}

Figure \ref{fig:seg_pics} presents some representative examples of segmentation model predictions. Each sub-image displays five B-scans selected from the 60 central B-scans. The predicted segmentation was generated using the sub-fold model corresponding to the test set to which the patient belongs. 

Figure \ref{fig:seg_pics1} illustrates a representative NPDR patient exhibiting severe intraretinal fluid. All four models successfully segment the majority of fluid regions. However, U-Net and SegFormer demonstrate weaker fine-layer segmentation performance than SwinUNETR and VM-UNet, particularly in the RNFL and OPL layers. VM-UNet excels in preserving layer continuity and structural integrity, whereas U-Net and SwinUNETR exhibit discontinuities in the OPL layer in the fourth and fifth B-scans. 

Figure \ref{fig:seg_pics2} displays a representative NPDR patient with pronounced HRF. Shading artifacts beneath large HRF clusters disrupt the continuity of the lower layers. VM-UNet demonstrates superior performance in maintaining layer integrity despite losing pixel intensity in the bottom three B-scans. In contrast, U-Net and SegFormer struggle to compensate for these artifacts, while SwinUNETR erroneously misclassifies portions of HRF within the Choroidal region. Notably, VM-UNet tends to under-segment the HRF relative to the other models.

Figure \ref{fig:seg_pics3} depicts a representative PDR patient with severe fluid accumulation. U-Net exhibits the weakest performance in fluid segmentation among all models, particularly in the third and fourth B-scans. Additionally, all models show varying degrees of under-segmentation in the final B-scan. 

Figure \ref{fig:seg_pics4} illustrates a representative NPDR patient with SRF. SwinUNETR and VM-UNet demonstrate superior SRF segmentation and refined surrounding layer boundaries compared to U-Net and SegFormer. SwinUNETR tends to over-segment both layers and fluid than VM-UNet in both layers and fluid, which is explicitly shown in the OPL layer and SRF of the fourth row. 

\begin{figure}[htb]
\centering
\includegraphics[width=\linewidth, keepaspectratio]{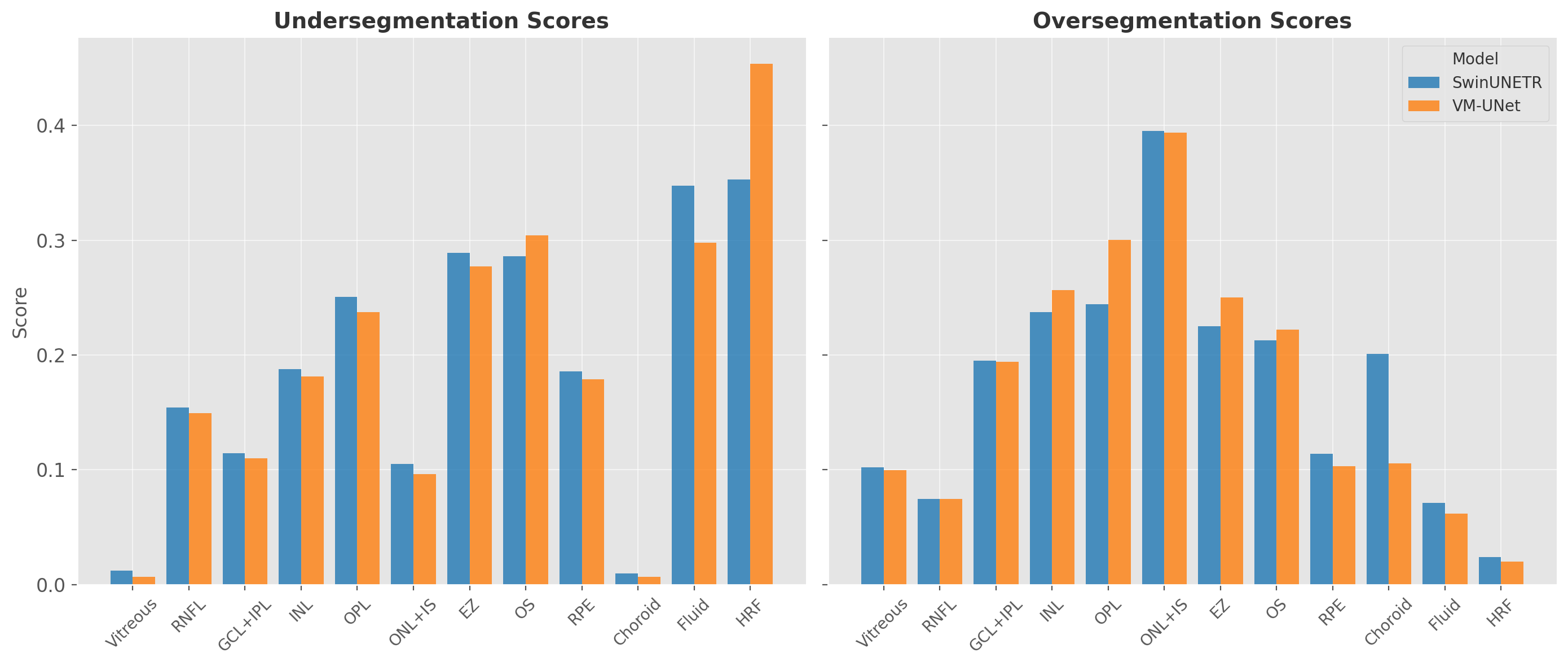}
\caption{Comparison of USS(left) and OSS(right) for SwinUNETR and VM-UNet. Lower values indicate better segmentation performance. A heuristic cutoff value of 0.2 was used to determine if there are under-segmentations or over-segmentations.}
\label{fig:seg_usos}
\end{figure}

Figure \ref{fig:seg_usos} shows the USS and OSS for the top-2 performance models SwinUNETR and VM-UNet. VM-UNet shows lower USS than SwinUNETR in most regions, except for HRF, and it has less over-segmentation in most areas, except for INL, OPL, EZ, and OS. Overall, using the 0.2 cutoff value, both models tend to under-segment in OPL, EZ and OS layers, plus the fluid and HRF. Over-segmentation is observed in the INL, OPL, ONL+IS, EZ and OS layers.

\subsection{Thickness}\label{resultthickness}

\begin{figure*}[htbp]
    \centering
    \includegraphics[height=0.65\textheight, keepaspectratio]{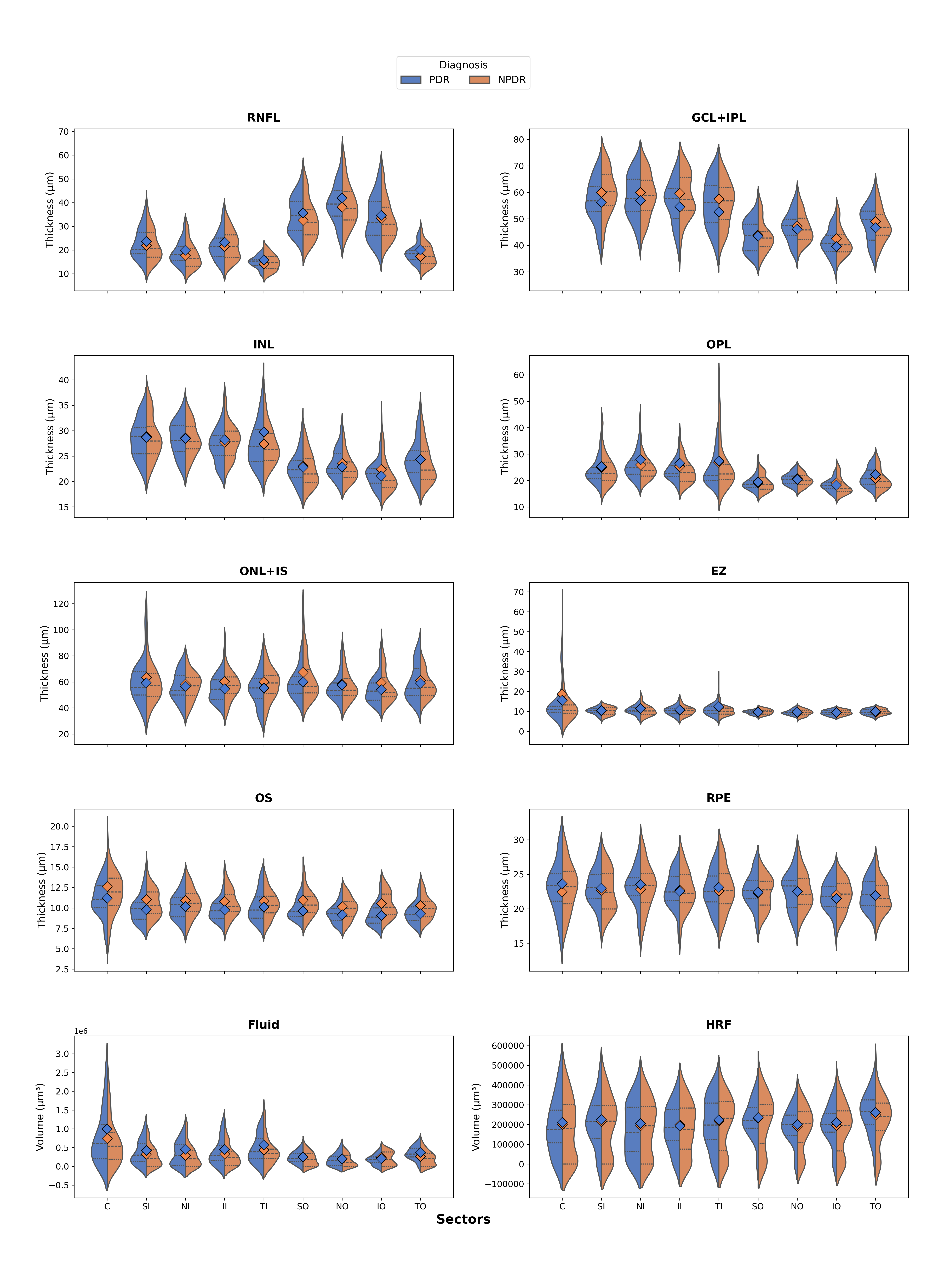}
    \caption{The distribution of layer thickness and fluid volume measurements across different sectors for patients diagnosed with NPDR and PDR. The outliers were removed outside the $5^{th}$ and $95^{th}$ percentiles. The diamond markers represent the mean thickness for each group, calculated from the original data without outlier removal. The central region thicknesses for inner retinal layers are not shown as mentioned in Section \ref{datathickness}.}
    \label{fig:thickness_plot}
\end{figure*}

Figure \ref{fig:thickness_plot} shows the violin plot of the distribution of the ground-truth-derived thickness across different retinal layers and sectors for NPDR and PDR groups. The outliers are removed outside the $5^{th}$ and $95^{th}$ percentiles, allowing a more robust interpretation of group differences. The diamond markers in each subplot show the mean thickness of each DR group without outlier removal. The PDR group has larger mean values and broader distributions in most sectors of RNFL, fluid and HRF, whereas the NPDR group has larger mean thickness in most sectors of GCL+IPL, ONL+IS and OS.

\begin{figure*}[htb]
\centering
\includegraphics[height=0.25\textheight, keepaspectratio]{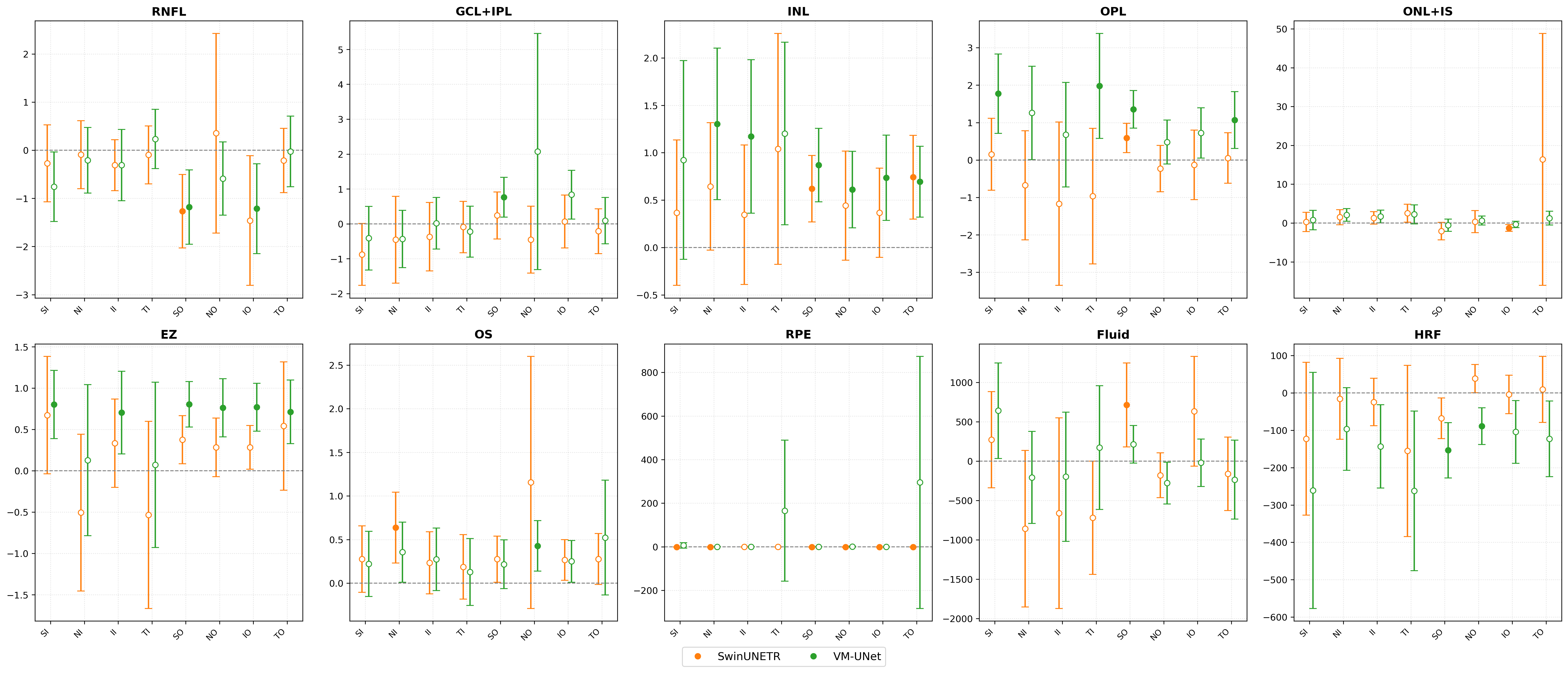}
\caption{Coefficient plot showing the statistical difference of thickness measurements between the model-predicted segmentation and ground truth based on GLM. The horizontal dashed line represents no difference in retinal thickness. The data with p-value $<$ 0.05 after FDR correction is annotated as filled markers. The data with p-value $>$ 0.05 after FDR correction but not across the reference line is marked as '*'. Error bars represent 95\% confidence intervals.}
\label{fig:thickness_comparison}
\end{figure*}

Figure \ref{fig:thickness_comparison} presents the deviations in retinal thickness measurements from SwinUNETR and VM-UNet segmentations compared to ground truth using GLM. Statistical significance is determined after controlling for multiple comparisons using FDR correction, with filled markers indicating FDR-adjusted p-values below 0.05. SwinUNETR generally demonstrates high agreement with ground truth, with minimal significant deviations except in the RPE layer, particularly in the SI, NI, SO, NO, IO, and TO sectors. In contrast, VM-UNet exhibits more widespread discrepancies, notably in the INL (NI, II, SO, NO, IO, TO), OPL (SI, TI, SO, TO), and EZ (SI, II, SO, NO, IO, TO) layers. Both models yield consistent predictions for fluid and HRF volumes. However, SwinUNETR shows significant overestimation in the SO sector of fluid, and VM-UNet displays significant under-segmentation in the SO and NO sectors of HRF. Additionally, abnormally large CIs are observed in specific sectors, including ONL+IS (TO) and OS (NO) for SwinUNETR, and GCL+IPL (NO) and RPE (TI, TO) for VM-UNet, likely reflecting segmentation failures that lead to extreme thickness estimates.

\begin{figure*}[htbp]
    \centering
    \includegraphics[height=0.25\textheight, keepaspectratio]{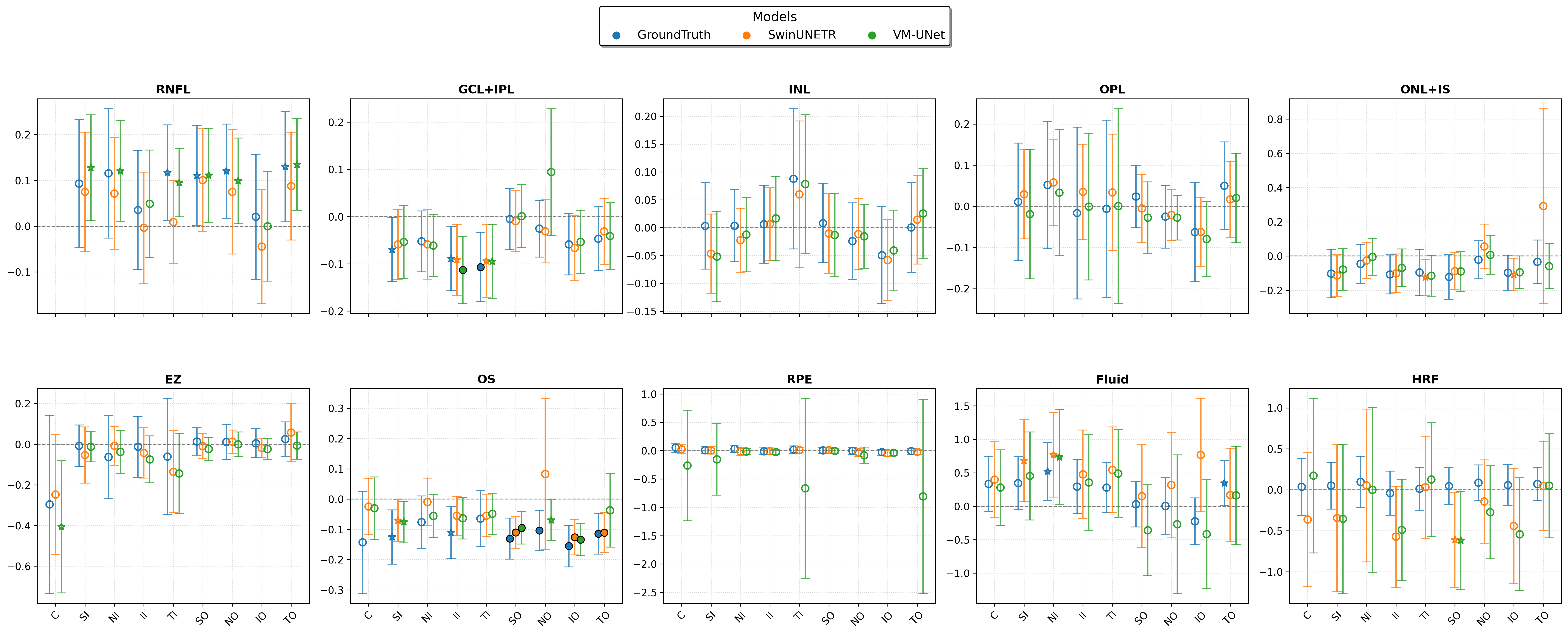}
    \caption{Coefficient plot illustrating the association between NPDR and PDR groups and retinal layer thickness across different retinal sectors, controlling for age, gender, and duration of diabetes. Results were derived from GLM analysis for three segmentation results: GroundTruth (blue circles), SwinUNETR (orange circles), and VM-UNet (green circles). The horizontal dashed line represents zero effect of DR diagnosis on retinal thickness. Open markers indicate non-significant associations. Markers labelled with an asterisk (*) represent data with p-values $> 0.05$ after FDR correction but not across the reference line. Error bars represent 95\% confidence intervals.}
    \label{fig:thickness_coefficient_plot}
\end{figure*}

Figure \ref{fig:thickness_coefficient_plot} presents statistical comparisons of retinal layer thickness, fluid volume, and HRF volume between NPDR and PDR groups. The analysis incorporates predicted segmentations from the two top-performing models, SwinUNETR and VM-UNet, alongside ground-truth segmentations for benchmarking. For each layer-sector pair, the GLM was applied to assess the relationship between DR diagnosis and thickness, accounting for potential confounders such as age, gender, and diabetes duration. A positive regression coefficient indicates increased thickness in the PDR group relative to NPDR.

Ground-truth data reveal a significantly increased thickness in multiple sectors (TI, SO, NO, TO) of the RNFL in PDR. Additionally, significant fluid accumulation was observed in the NI and TO sectors, respectively. Conversely, multiple sectors in GCL+IPL (SI, II, TI) and OS (SI, II, SO, NO, IO, TO) layers demonstrated significantly greater thickness in the NPDR group. After applying FDR correction, thinning in the TI sector of GCL+IPL and multiple sectors (SO, NO, IO, TO) of OS remain significant in the PDR group. Both segmentation models exhibited coefficient distributions consistent in most regions with the ground truth, especially for the significant layer sectors after FDR correction. Quantitative results are shown in Supplementary Table \ref{tab:thickness_glm}

\begin{figure*}[htbp]
    \centering

    \begin{subfigure}[b]{\textwidth}
        \centering
        \includegraphics[height=0.25\textheight, keepaspectratio]{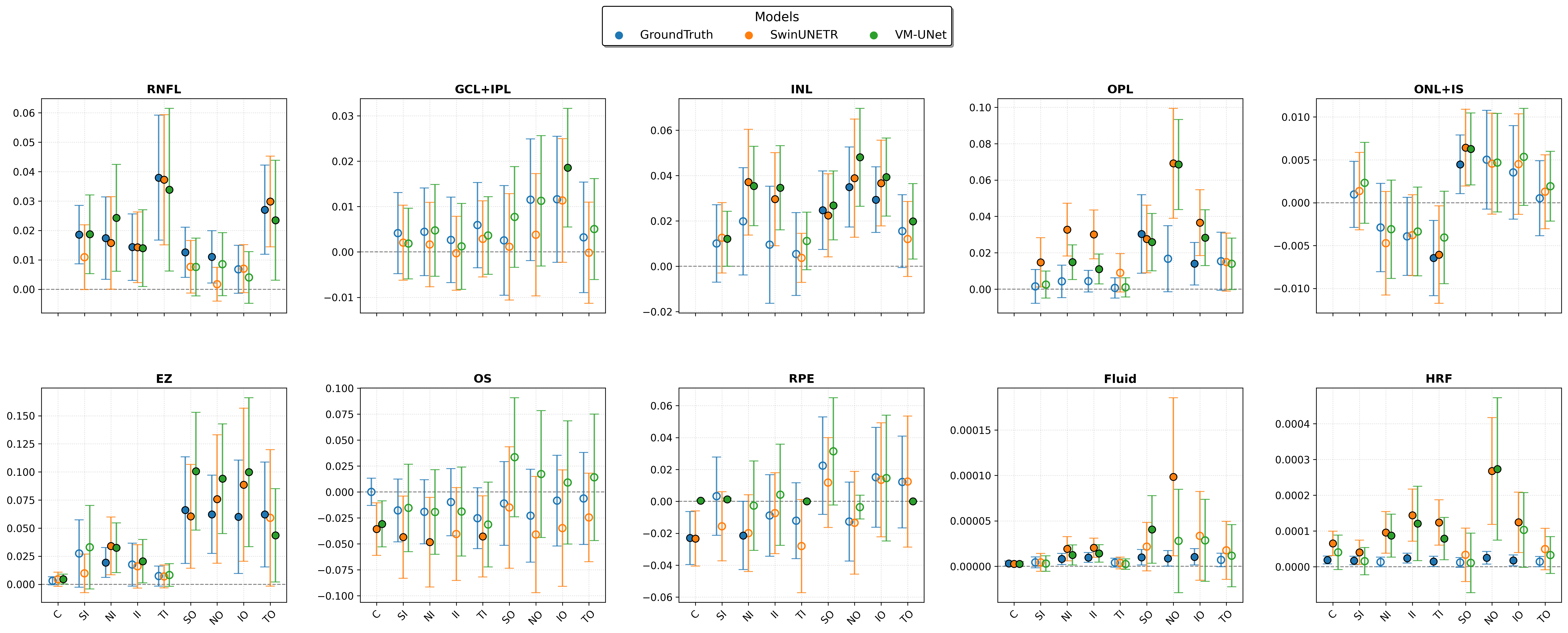}
        \subcaption{NPDR}
        \label{fig:VA_NPDR}
    \end{subfigure}
    \begin{subfigure}[b]{\textwidth}
        \centering
        \includegraphics[height=0.25\textheight, keepaspectratio]{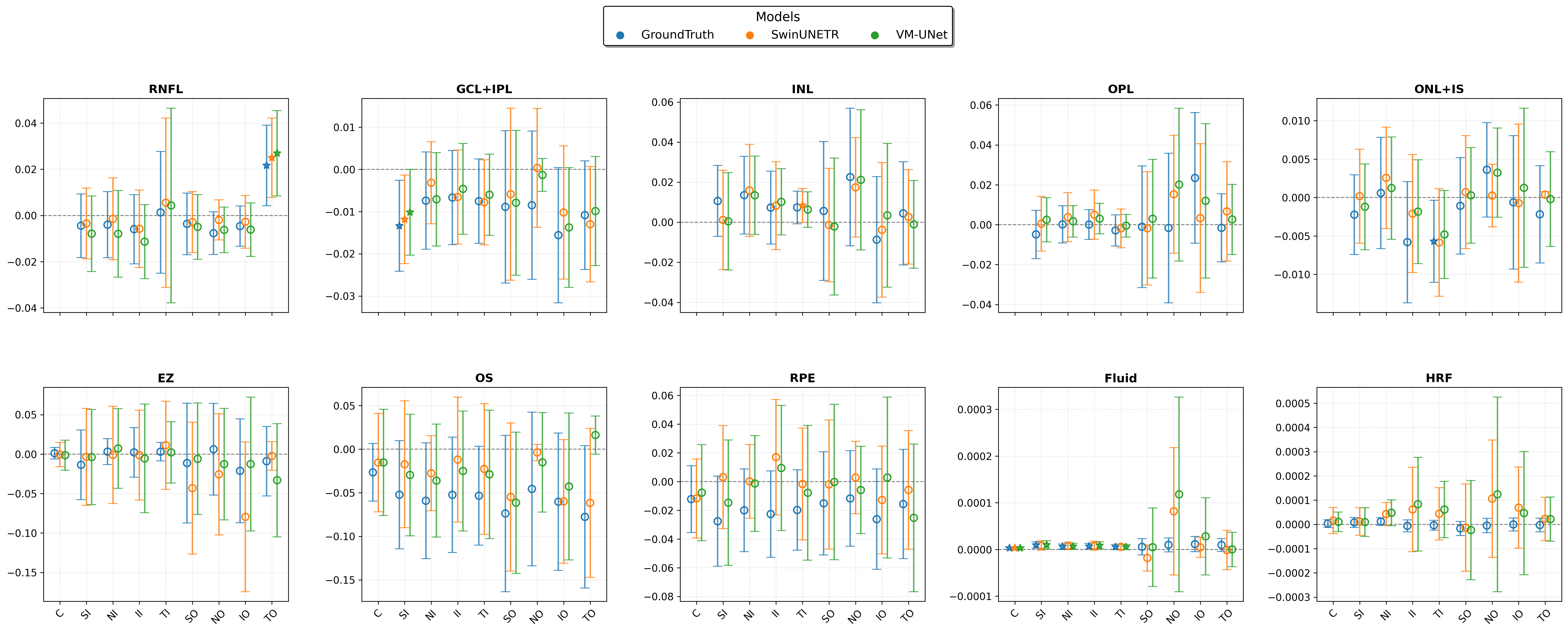}
        \subcaption{PDR}
        \label{fig:VA_PDR}
    \end{subfigure}
    \caption{Coefficient plots illustrating the association between the retinal layer thickness across different retinal sectors and visual acuity (logMAR), controlling for age, gender, and duration of diabetes. The analysis was performed for NPDR shown in \ref{fig:VA_NPDR} and PDR shown in \ref{fig:VA_PDR}, respectively. Results were derived from generalized linear model regression analyses for three segmentation results: GroundTruth (blue circles), SwinUNETR (orange circles), and VM-UNet (green circles). The horizontal dashed line represents no effect of retinal thickness on visual acuity. Error bars represent 95\% confidence intervals. Open markers indicate non-significant associations. Markers labelled with an asterisk (*) represent data with p-values $> 0.05$ after FDR correction but not across the reference line.  
    }
    \label{fig:VA}
\end{figure*}

Figure \ref{fig:etdrs_examples} presents four examples of thickness comparisons between NPDR and PDR groups using the ETDRS diagram described in Figure \ref{fig:etdrs}. Each example shows four regions that are reported with significant thickness differences in Figure \ref{fig:thickness_coefficient_plot}. Figures \ref{fig:etdrs1}–\ref{fig:etdrs4} correspond to cases of a 54-year-old female OS, a 44-year-old male OS, a 57-year-old male OD and a 53-year-old male OD, respectively. Each pair of patients is matched by age, gender, and eye laterality. For each retinal region, the first row displays the En Face image overlaid with the corresponding layer thickness heatmap, while the second row presents the sector-wise quantitative average thickness. The En Face image is generated using each layer's maximum intensity projection (MIP). For fluid, the En Face projection is derived from the entire retinal body (from the ILM to the BM), with thickness representing the accumulated volume in $\mu m^3$. From these figures \ref{fig:etdrs1}–\ref{fig:etdrs4}, PDR exhibits a larger thickness than NPDR in most RNFL sectors. Conversely, the thickness of inner sectors of GCL+IPL and most sectors of OS is significantly smaller for PDR. PDR has a larger and broader distribution of fluid accumulation in most sectors than NPDR. The findings are consistent with previous results.

\begin{figure*}[htbp]
    \centering
    \captionsetup{justification=centering}

    \begin{subfigure}[b]{0.48\textwidth}
        \centering
        \includegraphics[width=\linewidth, keepaspectratio]{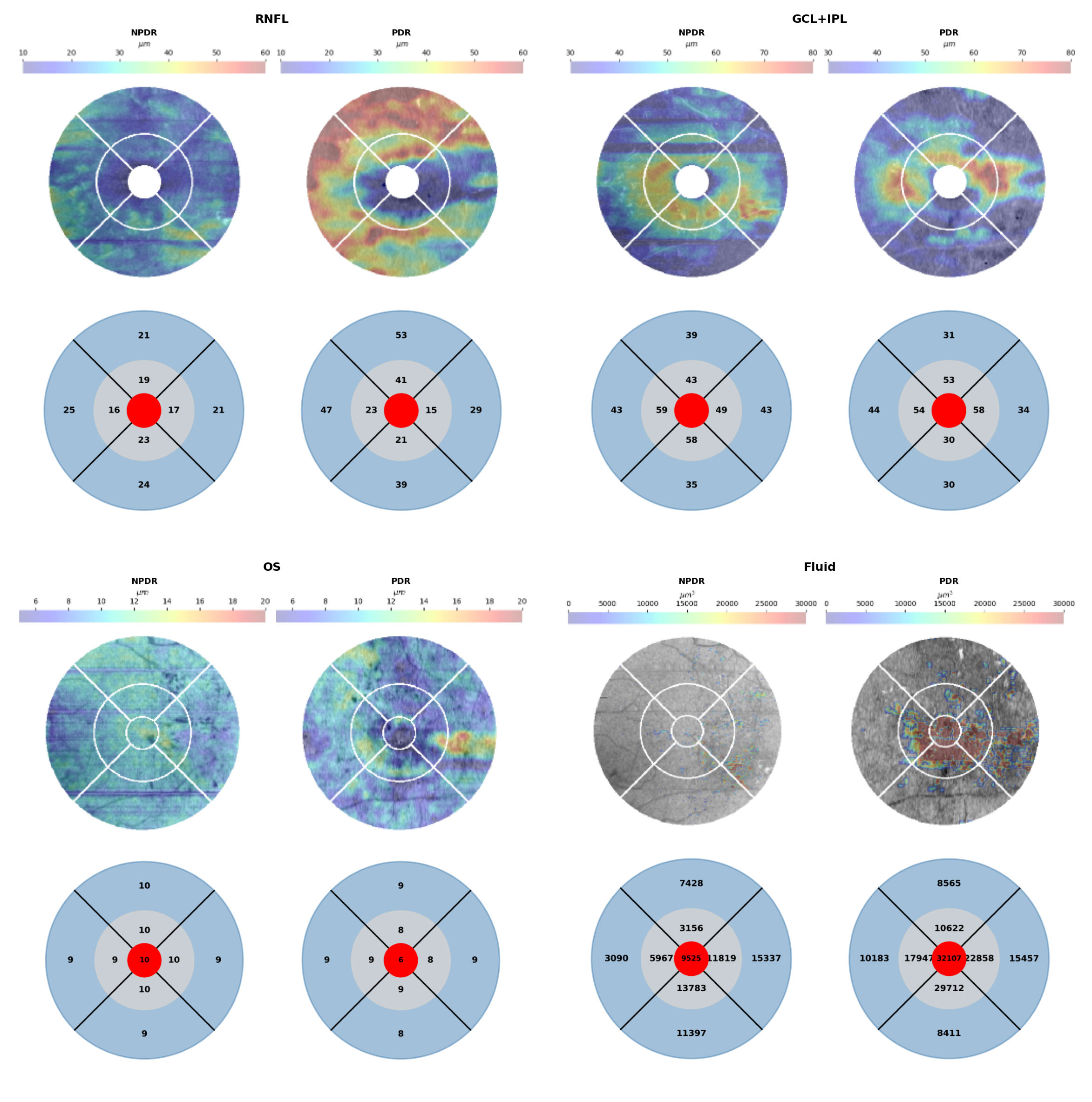}
        \subcaption{Example of a 54-year-old female patient with left eye image.}
        \label{fig:etdrs1}
    \end{subfigure}
    \begin{subfigure}[b]{0.48\textwidth}
        \centering
        \includegraphics[width=\linewidth, keepaspectratio]{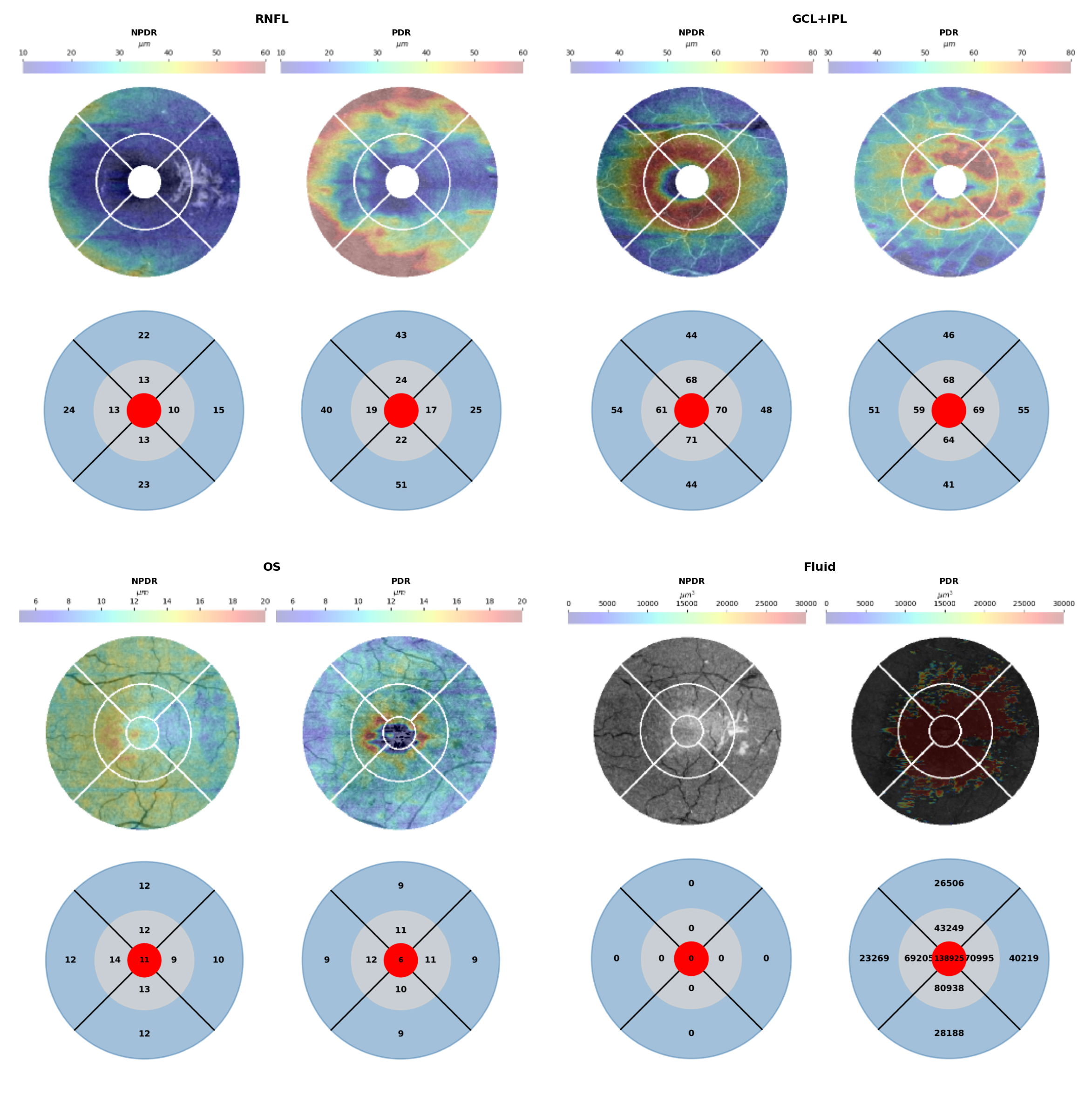}
        \subcaption{Example of a 44-year-old male patient with left eye image.}
        \label{fig:etdrs2}
    \end{subfigure}
\end{figure*}
\begin{figure*}[htbp]\ContinuedFloat-
    \centering
    
    \begin{subfigure}[b]{0.48\textwidth}
        \centering
        \includegraphics[width=\linewidth, keepaspectratio]{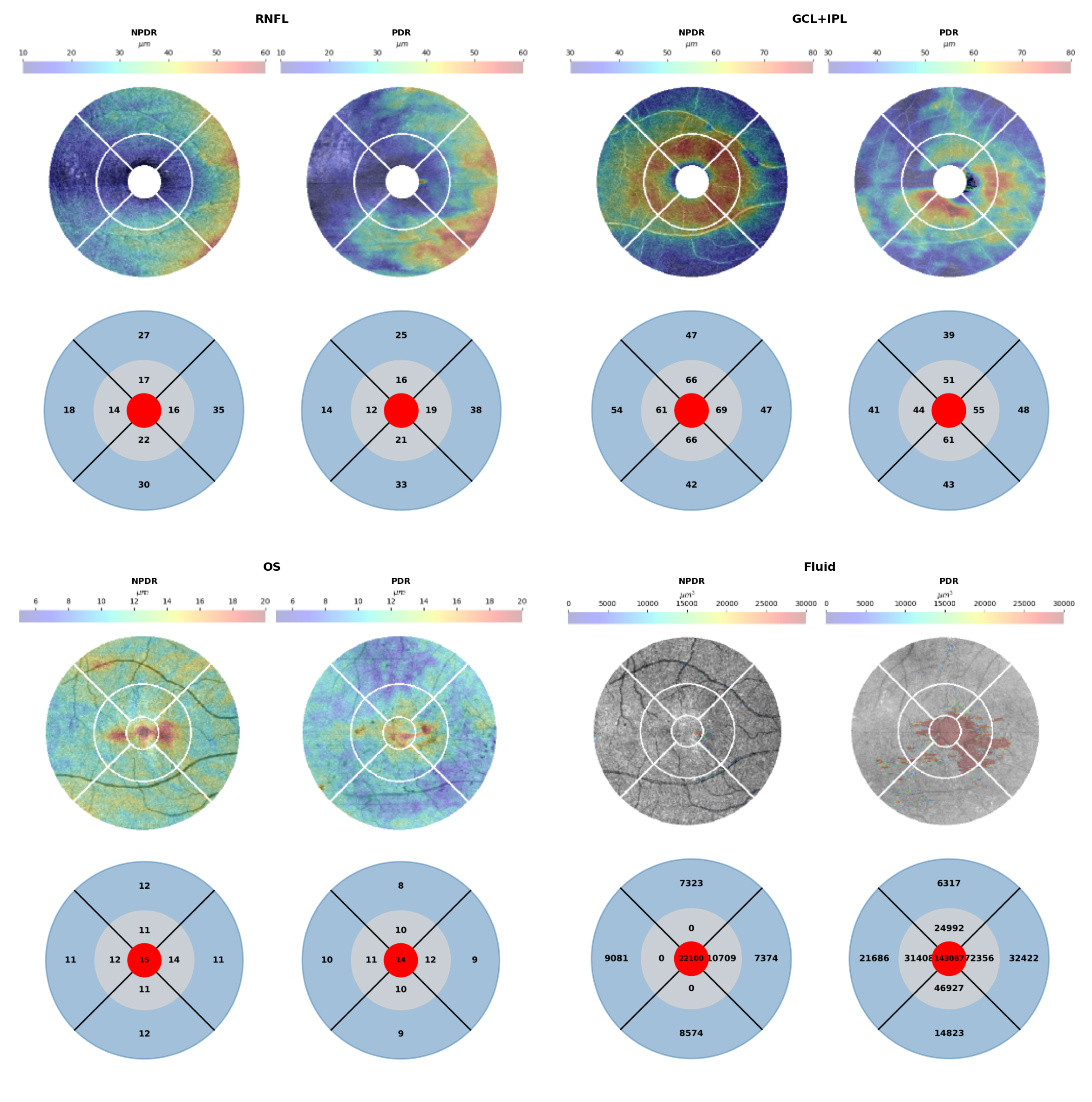}
        \subcaption{Example of a 57-year-old male patient with right eye image.}
        \label{fig:etdrs3}
    \end{subfigure}
    \begin{subfigure}[b]{0.48\textwidth}
        \centering
        \includegraphics[width=\linewidth, keepaspectratio]{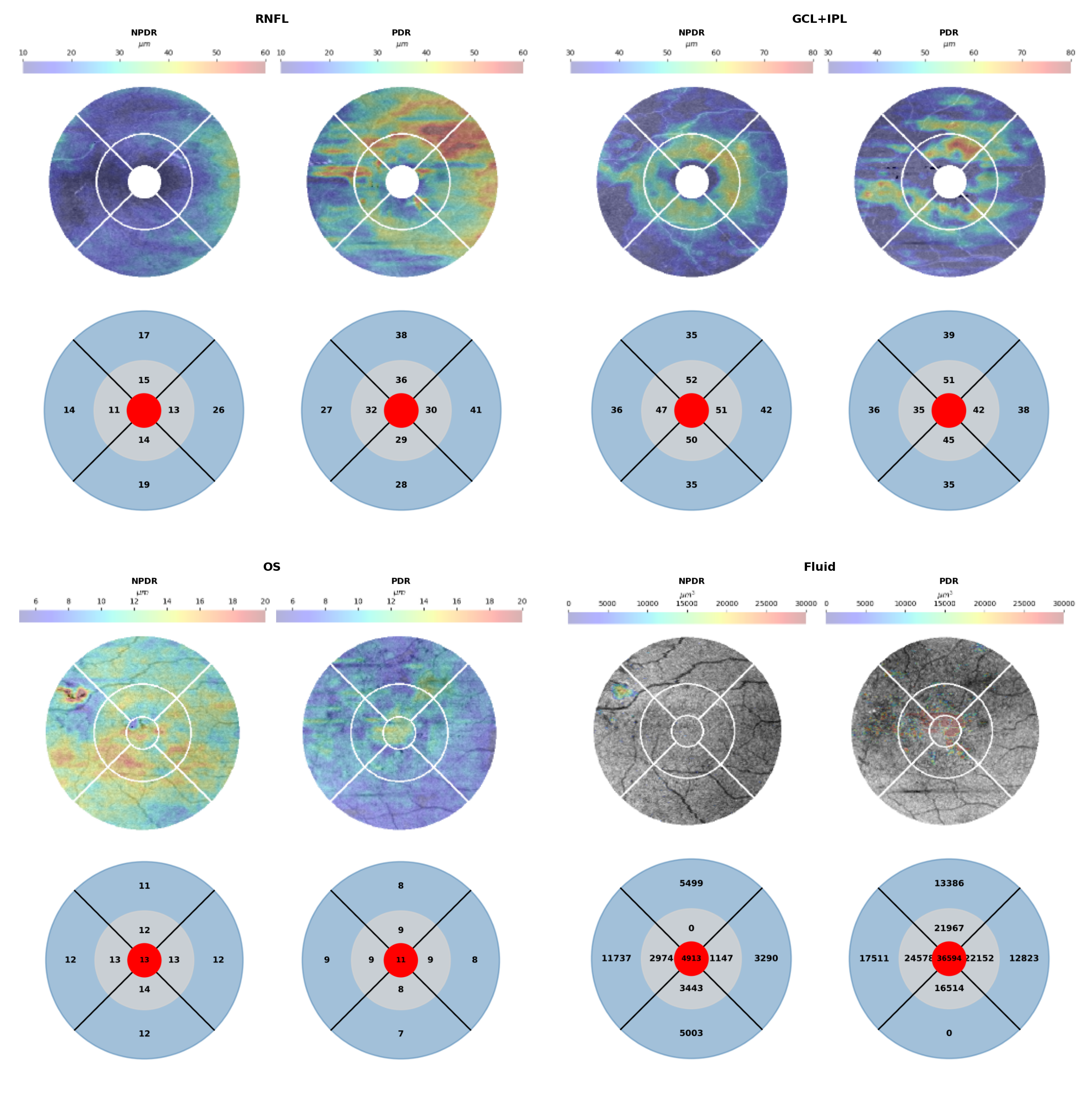}
        \subcaption{Example of a 53-year-old male patient with right eye image.}
        \label{fig:etdrs4}
    \end{subfigure}
    
    \caption{Examples of ETDRS thickness map comparison between NPDR and PDR patients with matched age, gender and eyeside. For each subfigure, only the four significant regions reported in Figure \ref{fig:thickness_coefficient_plot} are present. For each region, the NPDR and PDR groups are compared with two representations. The first row shows the thickness heatmap overlayed onto the layer En Face image. The second row shows the quantitative average layer thickness or volume accumulation for each sector.
    }
    \label{fig:etdrs_examples}

\end{figure*}

We further investigated the association between retinal layer-sector thickness and visual acuity (VA), with results summarized in Figure \ref{fig:VA}. This analysis was conducted separately for NPDR and PDR groups, using the same GLM regression framework applied in Figure \ref{fig:VA}, adjusting for age, gender, and diabetes duration. Quantitative results are shown in Supplementary Tables \ref{tab:VA_NPDR} and \ref{tab:VA_PDR}

In the NPDR group shown in Figure \ref{fig:VA_NPDR}, ground-truth segmentation revealed that the thickening of several layers was significantly associated with worse visual acuity (higher logMAR values). These included the RNFL (SI, NI, II, TI, SO, NO, TO), INL (SO, NO, IO), OPL (SO, IO), ONL+IS (SO), and EZ (NI, SO, NO, IO, TO). Layer thinning is significantly related to worse vision in ONL+IS (TI) and RPE (C, NI). Visual impairment was significantly correlated to fluid (C, NI, II, SSO, NO, IO) and HRF (C, SI, NI, II, TI, NO, IO) accumulation. All correlations remained significant after FDR correction. Both models predicted similar significant associations in most layer sectors, despite some false positives in OS for SwinUNETR and in RPE for VM-UNet. 

In the PDR group shown in Figure \ref{fig:VA_PDR}, the thickening of RNFL (TO) was significantly correlated with poorer vision. The thinning of the GCL+IPL (SI) and ONL+IS (TI) was also significantly correlated with worse vision. Fluid accumulation in multiple sectors (C, SI, NI, II, TI) was significantly correlated with vision loss, but no significant signs were shown for HRF. Both SwinUNETR and VM-UNet correctly identified most significant associations in RNFL, GCL+IPL and fluid regions. However, none of these associations remained statistically significant following FDR correction.

\section{Discussion}\label{discussion}
\subsection{Segmentation Models Comparison}\label{discussion_seg}
This study presents a comprehensive evaluation of the auto-segmentation performance with four state-of-the-art network architectures when segmenting retinal layer, fluid, and HRF on patients that exhibit varying levels of DR severity. The segmentation performance varied across models, highlighting differences in architectural strengths and their ability to segment specific retinal layers and fluid-related abnormalities. Specifically, SwinUNETR and VM-UNet consistently achieved high DSC and NSD scores, indicating their robustness in handling complex retinal structures. SwinUNETR particularly excelled in segmenting the OPL and HRF layers, which may be attributed to its transformer-based architecture that effectively captures long-range dependencies. VM-UNet, on the other hand, performed better in segmenting the fluid regions, suggesting that its sequential nature enhances segmentation continuity, particularly in areas with less distinct boundaries.

The performance differences in DSC and NSD indicate that while both models performed well, their strengths lie in different layers. VM-UNet was superior in several layers, plus fluid, whereas SwinUNETR demonstrated better performance in a few layers, plus HRF. U-Net and SegFormer, though competitive in some layers, exhibited weaker performance in fine layer segmentation, particularly in RNFL and OPL, where structural continuity is essential for accurate disease characterization. Although the SwinUNETR slightly outperforms VM-UNet in several regions, VM-UNet has significantly lower computational complexity (O(N)) than SwinUNETR (O($N^2$)), which is crucial for remote deployment in clinics with limited computational resources. 

The segmentation of fluid and HRF remains a significant challenge across all models. Fluid regions exhibit substantial variability, with VM-UNet demonstrating better spatial continuity but often under-segmenting these regions. In contrast, SwinUNETR captures fluid regions more extensively but is prone to occasional over-segmentation. HRF segmentation presents an even more significant challenge due to the presence of small, widely distributed hyper-reflective regions. Both models tend to under-segment fluid and HRF, frequently misclassifying them into adjacent retinal layers such as OPL and ONL+IS. Moreover, SwinUNETR generally exhibits a greater tendency to under-segment retinal regions than VM-UNet. The low DSC and NSD scores for fluid and HRF are mainly caused by their inherently small size relative to the full B-scan and their highly variable shapes and spatial distributions. For most scenarios, these structures occupy only a minor fraction of the retinal cross-section, making their accurate delineation more susceptible to minor boundary deviations. Additionally, their irregular morphology and variable positioning within the retina make consistent segmentation across patients particularly challenging, which disproportionately affects overlap-based metrics despite visually acceptable predictions. Although significant weight adjustments are applied to fluid and HRF regions, as described in Section \ref{datatrain}, additional strategies are needed to enhance model learning and improve segmentation performance in these complex regions.

The accuracy of the predicted segmentation was further evaluated through quantitative analysis of retinal layer thickness. Both SwinUNETR and VM-UNet demonstrated comparable performance, with minimal variation in thickness measurements across most layer-sector combinations when benchmarked against ground-truth annotations. Notably, in this cohort, SwinUNETR outperforms VM-UNet with superior consistency in thickness prediction across most layer sectors except the RPE. These findings suggest that while both models deliver comparable segmentation outputs, subtle segmentation inaccuracies can propagate non-linearly into downstream quantitative metrics such as thickness or volume. Such pixel-level deviations may become magnified in aggregate analyses, potentially leading to misinterpretation in studies relying on precise structural measurements.

\subsection{Clinical Insights}\label{discussion_clinical}
Significant differences in retinal layer thickness between NPDR and PDR offer valuable insights into the progression of DR. Ground-truth analysis revealed localized RNFL thickening in TI, SO, NO and TO sectors, GCL+IPL thinning in the SO sector and PS thinning in SO, NO, IO and TO sectors for PDR patients. This suggests a shift in the underlying disease mechanism—from inflammation-related thickening seen in NPDR to thinning caused by neurodegeneration and ischemia in PDR. Identifying disease stages through layer-specific biomarkers could support timely intervention. Therapies targeting edema or vascular leakage may be more beneficial during NPDR, while neuroprotective approaches might be essential for PDR curation. Similarly, increased fluid volume in the NI and TO sectors aligns with known patterns of retinal inflammation and exudation in advanced DR patients. Notably, both SwinUNETR and VM-UNet were able to replicate the general pattern of effect sizes seen in the ground-truth data, suggesting their suitability for detecting biologically meaningful trends despite minor segmentation discrepancies.

Our findings reveal significant associations between retinal layer thickness and visual acuity within NPDR and PDR patients. In the NPDR group, thickening of the RNFL in multiple sectors (SI, NI, II, TI, SO, NO, TO) may indicate early axoplasmic flow disruption and localized edema due to retinal ganglion cell dysfunction or vascular leakage. Similarly, thickening in the INL (SO, NO, IO) and OPL (SO, IO) likely reflects Müller cell swelling and intraretinal fluid accumulation—both hallmarks of early diabetic retinal changes. Thickening of the ONL+IS (SO) and EZ layers (NI, SO, NO, IO, TO) could represent subretinal fluid retention, inflammatory stress, or early disorganization of photoreceptor structures. Conversely, thinning of the ONL+IS (TI) and RPE (C, NI) was also associated with visual impairment, indicating focal photoreceptor loss or RPE atrophy. These degenerative changes impair phototransduction and outer retinal support, further degrading visual acuity. Additionally, accumulation of fluid (C, NI, II, SO, NO, IO) and HRF (C, SI, NI, II, TI, NO, IO) were significantly correlated with poor vision. These fluid-related biomarkers disrupt retinal architecture and light transmission, reinforcing the clinical relevance of monitoring retinal swelling and subcellular deposits in NPDR. 

In the PDR group, thickening of the RNFL in the TO sector was associated with reduced visual acuity, suggesting localized axonal swelling or edema in advanced stages. In contrast, thinning of the GCL+IPL (SI) and ONL+IS (TI) layers was also correlated with worse vision, indicative of progressive neurodegeneration and photoreceptor disruption. These findings imply the possible vascular leakage and retinal ischemia with neuronal loss. Additionally, fluid accumulation across multiple sectors (C, SI, NI, II, TI) showed significant associations with vision loss, consistent with the finding that macular edema is a major cause of visual decline. However, HRF accumulation did not demonstrate a significant association with visual acuity in this group, potentially reflecting a shift toward more diffuse or atrophic damage in late-stage disease. Nevertheless, none of these associations remained significant after FDR correction, highlighting the complex and heterogeneous retinal remodelling present in PDR. 

Both segmentation models effectively captured key associations between retinal layer thickness and visual acuity in NPDR and PDR patients, demonstrating strong alignment with ground-truth findings. They reliably identified structure–function relationships in clinically relevant layers such as RNFL, GCL+IPL, ONL+IS, and fluid. However, model-derived estimates occasionally showed greater variability, with wider confidence intervals likely due to segmentation noise or structural heterogeneity in advanced disease. Despite this, the results highlight the potential of deep learning models to predict vision impairment from retinal structural changes, supporting their use in automated risk assessment and clinical decision-making.

\section{Limitation and Future Work}
Despite the strengths of this study, several limitations must be acknowledged.

First, manual labelling imperfections impact both model performance and thickness measurements. Retinal layer segmentation is inherently challenging due to subtle boundary variations and overlapping structures. In cases where excessive fluid penetrates the layer boundaries, some portions of the layer become invisible or physically diminished. More clinical expertise is needed to segment the extreme instances properly. HRF segmentation suffers from inconsistencies in ground-truth annotations, as small, widely distributed foci are challenging to delineate manually. Interestingly, in some cases, automated models provided more precise segmentations than the ground-truth labels. For example, for the first B-scan in Figure \ref{fig:seg_pics1}, the predictions of SegFormer and SwinUNETR have better RNFL segmentation than the ground truth, with a smoother and more precise layer boundary. For the first B-scan in Figure \ref{fig:seg_pics3}, no fluid is manually annotated, but the segmentation models, except for SegFormer, predict potential intra-retinal fluid across the OPL and ONL+IS regions. Additionally, using more pre-processing and post-processing techniques may help improve the performance, such as the pixel-wise relative positional map as an extra input and a random forest classifier as a label refiner\cite{Ma2021LF-UNetImages}. 

Second, additional model comparisons may be necessary to provide a more comprehensive evaluation of segmentation approaches. While SwinUNETR and VM-UNet demonstrated superior performance, other architectures excel in certain perspectives. For example, MedSAM enables universal medical image segmentation with zero-shot capabilities\cite{Ma2023SegmentImages}. Zhu \textit{et al.} propose MedSAM-2 to perform medical segmentation tasks as a video object tracking problem\cite{Zhu2024Medical2}. The self-supervised few-shot semantic segmentation can be used for a limited number of labels\cite{Ouyang2022Self-SupervisedSegmentation}. However, many of these approaches rely on real-time user input or manual corrections, which diverges from the goal of this study—to assess whether widely adopted deep learning models can autonomously produce clinically reliable results. A broader comparison across multiple deep learning models could offer more insights into the trade-offs between performance, efficiency, and generalizability. The proposed model should also be validated from an external dataset, especially acquired from multiple OCT devices. To the best of our knowledge, there are currently no public datasets available with a comprehensive layer and fluid segmentation from severe DR patients. To advance benchmarking and promote the development of robust and generalizable segmentation models, greater efforts from the research community are needed to share diverse, well-annotated datasets. 

Third, this study's cross-sectional nature limits its ability to track disease progression over time. Longitudinal studies would provide better insights into how retinal layer thickness evolves in DR. For example, several studies report RNFL/GCL thinning during the progression of DR, which has become one of the most important preclinical biomarkers for DR severity evaluation\cite{Bhaskaran2023StudyTomography, Oshitari2009ChangesDiabetes., Park2011EarlyTomography.}. Additionally, while the sample size is sufficient to detect significant differences, it may limit the generalizability of the findings. A larger dataset encompassing a broader range of DR severities and treatment histories could provide more robust conclusions. In our study, no significant RNFL thickness difference is found between the NPDR and PDR groups. Fig \ref{fig:thickness_plot} shows explicit GCL+IPL thinning of PDR in terms of the mean and interquartile range, but only the NI sector exhibits marginal significance($p=0.058$). Expanding the cohort to include more diverse patient populations may help improve the applicability of the findings across different clinical settings.

Fourth, the lack of a detailed NPDR severity grading system may limit the ability to distinguish early, intermediate, and severe NPDR stages. Different NPDR severities likely exhibit distinct retinal layer changes, and a more granular classification system could enhance the understanding of DR progression. Future studies should explore integrating severity-based stratification to assess how thickness variations differ across NPDR subtypes. To our best knowledge, most DR grading datasets with public access focus on fundus color images like Messidor\footnote{https://www.adcis.net/en/third-party/messidor/} and DRTiD\footnote{https://github.com/FDU-VTS/DRTiD}. Additional efforts are needed to investigate the OCT image associated with DR severity levels, which are precisely determined using corresponding fundus images. 

Lastly, integrating multi-modal imaging techniques such as OCT angiography (OCTA) could provide additional insights into the vascular changes associated with DR. For example, Alam \textit{et al.} discovered the difference in vascular complexity features between NPDR and PDR patients\cite{Alam2021VASCULARRETINOPATHY}. Multiple OCT parameters are significantly correlated with DR severity\cite{Laotaweerungsawat2020OCTHospital}. Combining structural OCT findings with functional vascular imaging may improve disease characterization and facilitate more targeted therapeutic interventions. 

\section{Conclusion}
This study highlights the strengths and limitations of current deep learning-based segmentation models in analyzing diabetic retinopathy (DR)-related structural changes. Both SwinUNETR and VM-UNet exhibit strong performance, particularly in segmenting complex retinal layers and fluid regions. However, segmentation of fluid and HRF remains challenging due to their small size and dispersed distribution. Analysis of retinal layer thickness differences between NPDR and PDR reveals distinct structural alterations, with significant differences observed in the RNFL, GCL+IPL, OS, and fluid. The varying relationships between visual acuity and these structural changes in NPDR versus PDR suggest a progression from adaptive retinal remodelling in NPDR to pathological neurodegeneration and edema-driven vision loss in PDR, reinforcing the importance of early detection and intervention.

While the models enable detailed and efficient structural analysis, it is crucial to recognize that the choice of model can influence the clinical conclusions drawn from segmentation results. No single model consistently outperforms others across all tasks, highlighting the need to interpret findings in the context of model-specific strengths and weaknesses. The insights provided by these models contribute to our understanding of DR progression and may support improved disease classification and monitoring in clinical practice.

To further advance the clinical utility of automated OCT analysis, future work should address limitations such as manual labelling variability, the cross-sectional nature of the study, and the lack of fine-grained NPDR severity stratification. Incorporating longitudinal data, expanding the diversity and size of training datasets, and leveraging multi-modal imaging will benefit the robustness and predictive power of segmentation-based tools in DR treatment.


\section{Funding and acknowledgement}
We would like to thank Roy Boustani, Talha Mohammed and Shuting Xing for their help with manual segmentation and corrections. This study is funded by the National Sciences and Engineering
Research Council of Canada, Canadian Institutes of Health Research, Compute Canada, Wake Forest University School of Medicine Translational Eye and Vision Research (TREVR) Center, Moorfields Eye Charity and NIHR BRC at Moorfields and UCL IoO. The sponsor or funding organization had no role in the design or conduct of this research. The pipelines developed in this work will be available in our Cloud Engine Resource for Accelerated Medical Image Computing for Clinical Applications (CERAMICCA) platform (\url{https://ceramicca.from-ca.com/}).

\section{Declaration of competing interest}
All co-authors declare no conflict of interest.

\section{Declaration of generative AI and AI-assisted technologies in the writing process}

During the preparation of this work, the author(s) used ChatGPT (OpenAI) in order to improve the language and clarity of the manuscript. After using this tool, the author(s) reviewed and edited the content as needed and take(s) full responsibility for the content of the publication.


\bibliographystyle{IEEEtran}
\bibliography{SN_OCT.bib}

\newpage
\onecolumn
\section*{Supplementary Material}

\captionsetup{width=\textwidth}
\begin{landscape}
\input{ad_sd_significance_table}
\end{landscape}

\captionsetup{width=\textwidth}
\begin{landscape}
\input{latex_nested_glm_table}
\end{landscape}

\captionsetup{width=\textwidth}
\begin{landscape}
\input{va_glm_npdr_pdr_latex.tex}
\end{landscape}

\twocolumn 
\end{document}

%% file: ad_sd_significance_table.tex
\begin{table}
\centering
\scriptsize
\setlength{\tabcolsep}{2pt} 
\renewcommand{\arraystretch}{1.1}
\resizebox{\linewidth}{!}{
\begin{tabular}{|c|ccc|ccc|ccc|ccc|ccc|ccc|}
\hline
 & \multicolumn{3}{c|}{U-Net vs SegFormer} & \multicolumn{3}{c|}{U-Net vs SwinUNETR} & \multicolumn{3}{c|}{U-Net vs VM-UNet} & \multicolumn{3}{c|}{SegFormer vs SwinUNETR} & \multicolumn{3}{c|}{SegFormer vs VM-UNet} & \multicolumn{3}{c|}{SwinUNETR vs VM-UNet} \\
\cline{2-19}
 & $\bm{{\beta}}$ & p\_raw & p\_fdr & $\bm{{\beta}}$ & p\_raw & p\_fdr & $\bm{{\beta}}$ & p\_raw & p\_fdr & $\bm{{\beta}}$ & p\_raw & p\_fdr & $\bm{{\beta}}$ & p\_raw & p\_fdr & $\bm{{\beta}}$ & p\_raw & p\_fdr \\ \hline
RNFL & -0.0 & 0.849 & 0.849 & -0.001 & 0.255 & 0.383 & 0.002 & 0.054 & 0.108 & -0.001 & 0.34 & 0.408 & 0.002 & \textbf{0.035} & 0.105 & 0.003 & \textbf{0.003} & \textbf{0.019} \\
GCL+IPL & 0.0 & 0.637 & 0.637 & 0.003 & \textbf{0.0} & \textbf{0.0} & 0.006 & \textbf{0.0} & \textbf{0.0} & 0.003 & \textbf{0.001} & \textbf{0.001} & 0.006 & \textbf{0.0} & \textbf{0.0} & 0.003 & \textbf{0.002} & \textbf{0.002} \\
INL & -0.002 & \textbf{0.018} & \textbf{0.022} & 0.008 & \textbf{0.0} & \textbf{0.0} & 0.009 & \textbf{0.0} & \textbf{0.0} & 0.01 & \textbf{0.0} & \textbf{0.0} & 0.011 & \textbf{0.0} & \textbf{0.0} & 0.001 & 0.297 & 0.297 \\
OPL & -0.003 & \textbf{0.002} & \textbf{0.003} & 0.007 & \textbf{0.0} & \textbf{0.0} & 0.005 & \textbf{0.0} & \textbf{0.0} & 0.011 & \textbf{0.0} & \textbf{0.0} & 0.009 & \textbf{0.0} & \textbf{0.0} & -0.002 & 0.089 & 0.089 \\
ONL+IS & -0.001 & 0.144 & 0.216 & -0.0 & 0.718 & 0.718 & 0.003 & \textbf{0.0} & \textbf{0.0} & 0.001 & 0.299 & 0.359 & 0.005 & \textbf{0.0} & \textbf{0.0} & 0.004 & \textbf{0.0} & \textbf{0.0} \\
EZ & -0.004 & \textbf{0.004} & \textbf{0.004} & -0.007 & \textbf{0.0} & \textbf{0.0} & -0.013 & \textbf{0.0} & \textbf{0.0} & -0.003 & \textbf{0.011} & \textbf{0.011} & -0.009 & \textbf{0.0} & \textbf{0.0} & -0.006 & \textbf{0.0} & \textbf{0.0} \\
OS & -0.008 & \textbf{0.0} & \textbf{0.0} & -0.006 & \textbf{0.0} & \textbf{0.0} & -0.016 & \textbf{0.0} & \textbf{0.0} & 0.001 & 0.324 & 0.324 & -0.008 & \textbf{0.0} & \textbf{0.0} & -0.009 & \textbf{0.0} & \textbf{0.0} \\
RPE & -0.011 & \textbf{0.0} & \textbf{0.0} & -0.007 & \textbf{0.0} & \textbf{0.0} & -0.011 & \textbf{0.0} & \textbf{0.0} & 0.004 & \textbf{0.0} & \textbf{0.001} & -0.001 & 0.514 & 0.514 & -0.004 & \textbf{0.0} & \textbf{0.0} \\
Fluid & -0.04 & \textbf{0.0} & \textbf{0.0} & 0.028 & \textbf{0.0} & \textbf{0.0} & 0.029 & \textbf{0.0} & \textbf{0.0} & 0.068 & \textbf{0.0} & \textbf{0.0} & 0.069 & \textbf{0.0} & \textbf{0.0} & 0.001 & 0.773 & 0.773 \\
HRF & -0.085 & \textbf{0.0} & \textbf{0.0} & 0.033 & \textbf{0.0} & \textbf{0.0} & 0.014 & \textbf{0.0} & \textbf{0.0} & 0.117 & \textbf{0.0} & \textbf{0.0} & 0.098 & \textbf{0.0} & \textbf{0.0} & -0.019 & \textbf{0.0} & \textbf{0.0} \\
\hline
\end{tabular}%
}
\caption{Mean Dice Similarity Coefficient (DSC) comparison for each model pair using a generalized linear model (GLM). For each model pair, the retinal region, effect size, and p-value before and after FDR correction are indicated. Significant p-values are highlighted in \textbf{bold}.}
\label{tab:seg_glm_DSC}
\end{table}

\begin{table}[htbp]
\centering
\scriptsize
\setlength{\tabcolsep}{2pt} 
\renewcommand{\arraystretch}{1.1}
\resizebox{\linewidth}{!}{
\begin{tabular}{|c|ccc|ccc|ccc|ccc|ccc|ccc|}
\hline
 & \multicolumn{3}{c|}{U-Net vs SegFormer} & \multicolumn{3}{c|}{U-Net vs SwinUNETR} & \multicolumn{3}{c|}{U-Net vs VM-UNet} & \multicolumn{3}{c|}{SegFormer vs SwinUNETR} & \multicolumn{3}{c|}{SegFormer vs VM-UNet} & \multicolumn{3}{c|}{SwinUNETR vs VM-UNet} \\
\cline{2-19}
 & $\bm{{\beta}}$ & p\_raw & p\_fdr & $\bm{{\beta}}$ & p\_raw & p\_fdr & $\bm{{\beta}}$ & p\_raw & p\_fdr & $\bm{{\beta}}$ & p\_raw & p\_fdr & $\bm{{\beta}}$ & p\_raw & p\_fdr & $\bm{{\beta}}$ & p\_raw & p\_fdr \\ \hline
RNFL & 0.003 & \textbf{0.019} & \textbf{0.038} & 0.003 & \textbf{0.015} & \textbf{0.038} & 0.004 & \textbf{0.001} & \textbf{0.009} & 0.0 & 0.877 & 0.877 & 0.001 & 0.389 & 0.584 & 0.001 & 0.494 & 0.593 \\
GCL+IPL & 0.005 & \textbf{0.0} & \textbf{0.0} & 0.009 & \textbf{0.0} & \textbf{0.0} & 0.008 & \textbf{0.0} & \textbf{0.0} & 0.004 & \textbf{0.001} & \textbf{0.001} & 0.003 & \textbf{0.023} & \textbf{0.028} & -0.001 & 0.297 & 0.297 \\
INL & 0.003 & \textbf{0.047} & 0.056 & 0.011 & \textbf{0.0} & \textbf{0.0} & 0.009 & \textbf{0.0} & \textbf{0.0} & 0.008 & \textbf{0.0} & \textbf{0.0} & 0.006 & \textbf{0.0} & \textbf{0.0} & -0.002 & 0.193 & 0.193 \\
OPL & 0.006 & \textbf{0.0} & \textbf{0.0} & 0.011 & \textbf{0.0} & \textbf{0.0} & 0.004 & \textbf{0.002} & \textbf{0.002} & 0.005 & \textbf{0.0} & \textbf{0.0} & -0.001 & 0.356 & 0.356 & -0.007 & \textbf{0.0} & \textbf{0.0} \\
ONL+IS & 0.002 & 0.137 & 0.165 & 0.008 & \textbf{0.0} & \textbf{0.0} & 0.004 & \textbf{0.006} & \textbf{0.008} & 0.006 & \textbf{0.0} & \textbf{0.0} & 0.002 & 0.188 & 0.188 & -0.004 & \textbf{0.003} & \textbf{0.007} \\
EZ & 0.001 & 0.657 & 0.657 & -0.003 & \textbf{0.034} & 0.102 & -0.001 & 0.363 & 0.435 & -0.003 & \textbf{0.011} & 0.063 & -0.002 & 0.176 & 0.333 & 0.001 & 0.222 & 0.333 \\
OS & -0.002 & 0.096 & 0.115 & -0.005 & \textbf{0.0} & \textbf{0.0} & -0.006 & \textbf{0.0} & \textbf{0.0} & -0.003 & \textbf{0.016} & \textbf{0.025} & -0.004 & \textbf{0.002} & \textbf{0.003} & -0.001 & 0.491 & 0.491 \\
RPE & -0.009 & \textbf{0.0} & \textbf{0.0} & -0.006 & \textbf{0.0} & \textbf{0.0} & -0.02 & \textbf{0.0} & \textbf{0.0} & 0.003 & 0.123 & 0.123 & -0.011 & \textbf{0.0} & \textbf{0.0} & -0.014 & \textbf{0.0} & \textbf{0.0} \\
Fluid & -0.059 & \textbf{0.0} & \textbf{0.0} & 0.034 & \textbf{0.0} & \textbf{0.0} & 0.037 & \textbf{0.0} & \textbf{0.0} & 0.093 & \textbf{0.0} & \textbf{0.0} & 0.097 & \textbf{0.0} & \textbf{0.0} & 0.004 & 0.19 & 0.19 \\
HRF & -0.108 & \textbf{0.0} & \textbf{0.0} & 0.036 & \textbf{0.0} & \textbf{0.0} & 0.023 & \textbf{0.0} & \textbf{0.0} & 0.144 & \textbf{0.0} & \textbf{0.0} & 0.131 & \textbf{0.0} & \textbf{0.0} & -0.013 & \textbf{0.0} & \textbf{0.0} \\
\hline
\end{tabular}%
}
\caption{Mean Normalized Surface Dice (NSD) comparison for each model pair using a generalized linear model (GLM). For each model pair, the retinal region, effect size, and p-value before and after FDR correction are indicated. Significant p-values are highlighted in \textbf{bold}.}
\label{tab:seg_glm_NSD}
\end{table}

%% file: latex_nested_glm_table.tex
\begin{table}
\centering
\scriptsize
\renewcommand{\arraystretch}{1.1}
\resizebox{\linewidth}{!}{
\begin{tabular}{|c|c|ccc|ccc|ccc|ccc|ccc|ccc|ccc|ccc|ccc|}
\hline
\multicolumn{2}{|c|}{} & \multicolumn{3}{c|}{C} & \multicolumn{3}{c|}{SI} & \multicolumn{3}{c|}{NI} & \multicolumn{3}{c|}{II} & \multicolumn{3}{c|}{TI} & \multicolumn{3}{c|}{SO} & \multicolumn{3}{c|}{NO} & \multicolumn{3}{c|}{IO} & \multicolumn{3}{c|}{TO} \\
\multicolumn{2}{|c|}{} & $\beta$ & p$_\mathrm{raw}$ & p$_\mathrm{FDR}$ & $\beta$ & p$_\mathrm{raw}$ & p$_\mathrm{FDR}$ & $\beta$ & p$_\mathrm{raw}$ & p$_\mathrm{FDR}$ & $\beta$ & p$_\mathrm{raw}$ & p$_\mathrm{FDR}$ & $\beta$ & p$_\mathrm{raw}$ & p$_\mathrm{FDR}$ & $\beta$ & p$_\mathrm{raw}$ & p$_\mathrm{FDR}$ & $\beta$ & p$_\mathrm{raw}$ & p$_\mathrm{FDR}$ & $\beta$ & p$_\mathrm{raw}$ & p$_\mathrm{FDR}$ & $\beta$ & p$_\mathrm{raw}$ & p$_\mathrm{FDR}$ \\
\hline
\multirow{3}{*}{RNFL} & GT & / & / & / & 0.093 & 0.192 & 0.485 & 0.115 & 0.110 & 0.393 & 0.035 & 0.596 & 0.838 & 0.117 & \textbf{0.028} & 0.263 & 0.110 & \textbf{0.047} & 0.279 & 0.120 & \textbf{0.022} & 0.222 & 0.020 & 0.770 & 0.889 & 0.130 & \textbf{0.034} & 0.264 \\
 & Swin & / & / & / & 0.075 & 0.263 & 0.588 & 0.071 & 0.251 & 0.575 & -0.004 & 0.954 & 0.949 & 0.009 & 0.847 & 0.923 & 0.101 & 0.079 & 0.345 & 0.075 & 0.280 & 0.601 & -0.045 & 0.483 & 0.784 & 0.088 & 0.145 & 0.441 \\
 & VM & / & / & / & 0.127 & \textbf{0.031} & 0.263 & 0.120 & \textbf{0.032} & 0.263 & 0.049 & 0.416 & 0.732 & 0.095 & \textbf{0.013} & 0.197 & 0.111 & \textbf{0.034} & 0.264 & 0.099 & \textbf{0.039} & 0.277 & -0.000 & 0.995 & 0.960 & 0.135 & \textbf{0.008} & 0.156 \\

\hline
\multirow{3}{*}{GCL+IPL} & GT & / & / & / & -0.070 & \textbf{0.045} & 0.277 & -0.052 & 0.111 & 0.393 & -0.089 & \textbf{0.010} & 0.177 & -0.107 & \textbf{0.004} & 0.096 & -0.005 & 0.877 & 0.923 & -0.025 & 0.408 & 0.729 & -0.059 & 0.076 & 0.345 & -0.047 & 0.178 & 0.465 \\
 & Swin & / & / & / & -0.059 & 0.122 & 0.393 & -0.059 & 0.117 & 0.393 & -0.092 & \textbf{0.017} & 0.206 & -0.094 & \textbf{0.017} & 0.206 & -0.010 & 0.773 & 0.889 & -0.031 & 0.362 & 0.677 & -0.066 & 0.058 & 0.326 & -0.031 & 0.378 & 0.692 \\
 & VM & / & / & / & -0.054 & 0.172 & 0.460 & -0.061 & 0.065 & 0.328 & -0.113 & \textbf{0.002} & 0.052 & -0.095 & \textbf{0.018} & 0.206 & 0.001 & 0.979 & 0.960 & 0.094 & 0.169 & 0.460 & -0.053 & 0.115 & 0.393 & -0.041 & 0.254 & 0.577 \\

\hline
\multirow{3}{*}{INL} & GT & / & / & / & 0.003 & 0.939 & 0.939 & 0.003 & 0.921 & 0.932 & 0.006 & 0.865 & 0.923 & 0.088 & 0.173 & 0.460 & 0.008 & 0.822 & 0.923 & -0.024 & 0.488 & 0.786 & -0.050 & 0.264 & 0.588 & 0.000 & 0.995 & 0.960 \\
 & Swin & / & / & / & -0.047 & 0.198 & 0.494 & -0.023 & 0.439 & 0.742 & 0.006 & 0.849 & 0.923 & 0.060 & 0.374 & 0.692 & -0.011 & 0.771 & 0.889 & -0.012 & 0.722 & 0.870 & -0.058 & 0.116 & 0.393 & 0.014 & 0.725 & 0.870 \\
 & VM & / & / & / & -0.052 & 0.208 & 0.500 & -0.012 & 0.717 & 0.870 & 0.017 & 0.667 & 0.870 & 0.078 & 0.219 & 0.508 & -0.013 & 0.728 & 0.870 & -0.016 & 0.594 & 0.838 & -0.041 & 0.269 & 0.592 & 0.025 & 0.536 & 0.817 \\

\hline
\multirow{3}{*}{OPL} & GT & / & / & / & 0.011 & 0.884 & 0.925 & 0.052 & 0.511 & 0.802 & -0.016 & 0.878 & 0.923 & -0.006 & 0.956 & 0.949 & 0.024 & 0.537 & 0.817 & -0.025 & 0.519 & 0.804 & -0.063 & 0.306 & 0.641 & 0.050 & 0.358 & 0.676 \\
 & Swin & / & / & / & 0.029 & 0.595 & 0.838 & 0.058 & 0.279 & 0.601 & 0.035 & 0.555 & 0.824 & 0.034 & 0.641 & 0.863 & -0.005 & 0.903 & 0.932 & -0.021 & 0.498 & 0.792 & -0.062 & 0.145 & 0.441 & 0.017 & 0.723 & 0.870 \\
 & VM & / & / & / & -0.019 & 0.813 & 0.923 & 0.033 & 0.670 & 0.870 & -0.001 & 0.993 & 0.960 & 0.000 & 0.997 & 0.960 & -0.028 & 0.533 & 0.817 & -0.028 & 0.319 & 0.651 & -0.080 & 0.084 & 0.349 & 0.020 & 0.713 & 0.870 \\

\hline
\multirow{3}{*}{ONL+IS} & GT & / & / & / & -0.103 & 0.156 & 0.456 & -0.046 & 0.430 & 0.738 & -0.107 & 0.066 & 0.328 & -0.096 & 0.165 & 0.460 & -0.122 & 0.067 & 0.328 & -0.021 & 0.707 & 0.870 & -0.097 & 0.065 & 0.328 & -0.034 & 0.605 & 0.838 \\
 & Swin & / & / & / & -0.113 & 0.068 & 0.328 & -0.026 & 0.630 & 0.862 & -0.101 & 0.079 & 0.345 & -0.124 & \textbf{0.020} & 0.214 & -0.088 & 0.111 & 0.393 & 0.056 & 0.403 & 0.726 & -0.108 & \textbf{0.027} & 0.263 & 0.292 & 0.316 & 0.650 \\
 & VM & / & / & / & -0.079 & 0.204 & 0.495 & -0.005 & 0.934 & 0.937 & -0.069 & 0.220 & 0.508 & -0.115 & 0.059 & 0.326 & -0.090 & 0.131 & 0.416 & 0.007 & 0.909 & 0.932 & -0.095 & 0.051 & 0.298 & -0.059 & 0.378 & 0.692 \\

\hline
\multirow{3}{*}{EZ} & GT & -0.296 & 0.185 & 0.477 & -0.008 & 0.875 & 0.923 & -0.063 & 0.544 & 0.822 & -0.013 & 0.867 & 0.923 & -0.061 & 0.676 & 0.870 & 0.013 & 0.704 & 0.870 & 0.010 & 0.816 & 0.923 & 0.004 & 0.904 & 0.932 & 0.025 & 0.569 & 0.828 \\
 & Swin & -0.248 & 0.098 & 0.375 & -0.053 & 0.450 & 0.750 & -0.008 & 0.873 & 0.923 & -0.043 & 0.491 & 0.786 & -0.136 & 0.187 & 0.478 & -0.010 & 0.748 & 0.881 & 0.012 & 0.671 & 0.870 & -0.018 & 0.459 & 0.754 & 0.057 & 0.429 & 0.738 \\
 & VM & -0.406 & \textbf{0.014} & 0.206 & -0.013 & 0.741 & 0.881 & -0.038 & 0.482 & 0.784 & -0.075 & 0.202 & 0.495 & -0.144 & 0.152 & 0.453 & -0.023 & 0.428 & 0.738 & -0.001 & 0.983 & 0.960 & -0.024 & 0.355 & 0.676 & -0.007 & 0.834 & 0.923 \\

\hline
\multirow{3}{*}{OS} & GT & -0.143 & 0.098 & 0.375 & -0.125 & \textbf{0.006} & 0.124 & -0.076 & 0.083 & 0.349 & -0.111 & \textbf{0.011} & 0.181 & -0.065 & 0.172 & 0.460 & -0.131 & \textbf{0.000} & \textbf{0.008} & -0.103 & \textbf{0.002} & 0.059 & -0.155 & \textbf{0.000} & \textbf{0.001} & -0.115 & \textbf{0.001} & \textbf{0.029} \\
 & Swin & -0.025 & 0.603 & 0.838 & -0.070 & \textbf{0.045} & 0.277 & -0.009 & 0.815 & 0.923 & -0.055 & 0.095 & 0.375 & -0.055 & 0.117 & 0.393 & -0.110 & \textbf{0.000} & \textbf{0.002} & 0.083 & 0.516 & 0.804 & -0.126 & \textbf{0.000} & \textbf{0.002} & -0.111 & \textbf{0.001} & \textbf{0.029} \\
 & VM & -0.030 & 0.568 & 0.828 & -0.076 & \textbf{0.031} & 0.263 & -0.056 & 0.121 & 0.393 & -0.063 & 0.068 & 0.328 & -0.049 & 0.165 & 0.460 & -0.095 & \textbf{0.000} & \textbf{0.020} & -0.070 & \textbf{0.041} & 0.277 & -0.134 & \textbf{0.000} & \textbf{0.000} & -0.037 & 0.550 & 0.822 \\

\hline
\multirow{3}{*}{RPE} & GT & 0.052 & 0.219 & 0.508 & 0.007 & 0.830 & 0.923 & 0.031 & 0.351 & 0.676 & -0.011 & 0.719 & 0.870 & 0.019 & 0.568 & 0.828 & 0.004 & 0.871 & 0.923 & -0.003 & 0.925 & 0.932 & -0.026 & 0.336 & 0.664 & -0.009 & 0.745 & 0.881 \\
 & Swin & 0.023 & 0.571 & 0.828 & 0.004 & 0.895 & 0.932 & -0.016 & 0.654 & 0.870 & -0.009 & 0.773 & 0.889 & 0.010 & 0.706 & 0.870 & 0.011 & 0.686 & 0.870 & -0.032 & 0.359 & 0.676 & -0.046 & 0.084 & 0.349 & -0.022 & 0.328 & 0.660 \\
 & VM & -0.262 & 0.599 & 0.838 & -0.154 & 0.633 & 0.862 & -0.015 & 0.596 & 0.838 & -0.025 & 0.300 & 0.633 & -0.665 & 0.412 & 0.731 & -0.004 & 0.850 & 0.923 & -0.082 & 0.271 & 0.592 & -0.038 & 0.060 & 0.326 & -0.808 & 0.355 & 0.676 \\

\hline
\multirow{3}{*}{Fluid} & GT & 0.333 & 0.113 & 0.393 & 0.345 & 0.087 & 0.355 & 0.519 & \textbf{0.019} & 0.206 & 0.292 & 0.155 & 0.456 & 0.278 & 0.146 & 0.441 & 0.029 & 0.866 & 0.923 & 0.003 & 0.989 & 0.960 & -0.226 & 0.204 & 0.495 & 0.344 & \textbf{0.044} & 0.277 \\
 & Swin & 0.398 & 0.168 & 0.460 & 0.681 & \textbf{0.030} & 0.263 & 0.768 & \textbf{0.017} & 0.206 & 0.476 & 0.159 & 0.458 & 0.543 & 0.096 & 0.375 & 0.150 & 0.702 & 0.870 & 0.316 & 0.432 & 0.738 & 0.766 & 0.076 & 0.345 & 0.168 & 0.639 & 0.863 \\
 & VM & 0.280 & 0.330 & 0.660 & 0.453 & 0.177 & 0.465 & 0.734 & \textbf{0.042} & 0.277 & 0.355 & 0.332 & 0.660 & 0.488 & 0.143 & 0.441 & -0.360 & 0.299 & 0.633 & -0.269 & 0.610 & 0.840 & -0.417 & 0.315 & 0.650 & 0.163 & 0.665 & 0.870 \\

\hline
\multirow{3}{*}{HRF} & GT & 0.037 & 0.836 & 0.923 & 0.051 & 0.725 & 0.870 & 0.096 & 0.548 & 0.822 & -0.041 & 0.767 & 0.889 & 0.013 & 0.924 & 0.932 & 0.046 & 0.690 & 0.870 & 0.087 & 0.434 & 0.738 & 0.057 & 0.650 & 0.870 & 0.070 & 0.502 & 0.794 \\
 & Swin & -0.363 & 0.384 & 0.697 & -0.344 & 0.453 & 0.750 & 0.053 & 0.912 & 0.932 & -0.572 & 0.070 & 0.328 & 0.031 & 0.922 & 0.932 & -0.611 & \textbf{0.039} & 0.277 & -0.143 & 0.582 & 0.838 & -0.442 & 0.218 & 0.508 & 0.047 & 0.865 & 0.923 \\
 & VM & 0.173 & 0.720 & 0.870 & -0.355 & 0.445 & 0.746 & -0.000 & 0.999 & 0.960 & -0.490 & 0.121 & 0.393 & 0.125 & 0.724 & 0.870 & -0.617 & \textbf{0.043} & 0.277 & -0.273 & 0.346 & 0.676 & -0.544 & 0.122 & 0.393 & 0.050 & 0.877 & 0.923 \\

\hline
\end{tabular}%
}
\caption{Retinal layer thickness, fluid and HRF volume comparison for each model pair using generalized linear model (GLM) while controlling for age, gender and diabetes duration. For each model pair, the retinal layer, effect size, p-value before and after FDR correction are indicated.}
\label{tab:thickness_glm}
\end{table}

%% file: va_glm_npdr_pdr_latex.tex
\begin{table}[htbp]
\centering
\scriptsize
\renewcommand{\arraystretch}{1.1}
\resizebox{\linewidth}{!}{
\begin{tabular}{|c|c|ccc|ccc|ccc|ccc|ccc|ccc|ccc|ccc|ccc|}
\hline
\multicolumn{2}{|c|}{} & \multicolumn{3}{c|}{C} & \multicolumn{3}{c|}{SI} & \multicolumn{3}{c|}{NI} & \multicolumn{3}{c|}{II} & \multicolumn{3}{c|}{TI} & \multicolumn{3}{c|}{SO} & \multicolumn{3}{c|}{NO} & \multicolumn{3}{c|}{IO} & \multicolumn{3}{c|}{TO} \\
\multicolumn{2}{|c|}{} & $\beta$ & p$_\mathrm{raw}$ & p$_\mathrm{FDR}$ & $\beta$ & p$_\mathrm{raw}$ & p$_\mathrm{FDR}$ & $\beta$ & p$_\mathrm{raw}$ & p$_\mathrm{FDR}$ & $\beta$ & p$_\mathrm{raw}$ & p$_\mathrm{FDR}$ & $\beta$ & p$_\mathrm{raw}$ & p$_\mathrm{FDR}$ & $\beta$ & p$_\mathrm{raw}$ & p$_\mathrm{FDR}$ & $\beta$ & p$_\mathrm{raw}$ & p$_\mathrm{FDR}$ & $\beta$ & p$_\mathrm{raw}$ & p$_\mathrm{FDR}$ & $\beta$ & p$_\mathrm{raw}$ & p$_\mathrm{FDR}$ \\
\hline    
\multirow{3}{*}{RNFL} & GT & / & / & / & 0.019 & \textbf{0.000} & \textbf{0.002} & 0.017 & \textbf{0.015} & \textbf{0.030} & 0.014 & \textbf{0.013} & \textbf{0.027} & 0.038 & \textbf{0.000} & \textbf{0.003} & 0.013 & \textbf{0.004} & \textbf{0.014} & 0.011 & \textbf{0.015} & \textbf{0.030} & 0.007 & 0.103 & 0.110 & 0.027 & \textbf{0.000} & \textbf{0.003} \\
 & Swin & / & / & / & 0.011 & 0.052 & 0.072 & 0.016 & \textbf{0.050} & 0.070 & 0.014 & \textbf{0.020} & \textbf{0.036} & 0.037 & \textbf{0.001} & \textbf{0.005} & 0.008 & 0.095 & 0.106 & 0.002 & 0.560 & 0.381 & 0.007 & 0.093 & 0.106 & 0.030 & \textbf{0.000} & \textbf{0.002} \\
 & VM & / & / & / & 0.019 & \textbf{0.006} & \textbf{0.016} & 0.024 & \textbf{0.009} & \textbf{0.020} & 0.014 & \textbf{0.036} & 0.055 & 0.034 & \textbf{0.016} & \textbf{0.032} & 0.008 & 0.131 & 0.128 & 0.009 & 0.119 & 0.119 & 0.004 & 0.369 & 0.282 & 0.023 & \textbf{0.024} & \textbf{0.041} \\

\hline
\multirow{3}{*}{GCL+IPL} & GT & / & / & / & 0.004 & 0.363 & 0.280 & 0.004 & 0.367 & 0.282 & 0.003 & 0.582 & 0.387 & 0.006 & 0.216 & 0.189 & 0.003 & 0.680 & 0.437 & 0.012 & 0.092 & 0.106 & 0.012 & 0.102 & 0.110 & 0.003 & 0.605 & 0.399 \\
 & Swin & / & / & / & 0.002 & 0.628 & 0.409 & 0.002 & 0.730 & 0.463 & -0.000 & 0.943 & 0.570 & 0.003 & 0.501 & 0.353 & 0.001 & 0.849 & 0.516 & 0.004 & 0.582 & 0.387 & 0.011 & 0.103 & 0.110 & -0.000 & 0.975 & 0.588 \\
 & VM & / & / & / & 0.002 & 0.641 & 0.416 & 0.005 & 0.359 & 0.278 & 0.001 & 0.800 & 0.494 & 0.004 & 0.406 & 0.302 & 0.008 & 0.173 & 0.157 & 0.011 & 0.124 & 0.123 & 0.019 & \textbf{0.005} & \textbf{0.016} & 0.005 & 0.374 & 0.285 \\

\hline
\multirow{3}{*}{INL} & GT & / & / & / & 0.010 & 0.248 & 0.213 & 0.020 & 0.100 & 0.110 & 0.010 & 0.470 & 0.338 & 0.005 & 0.564 & 0.382 & 0.025 & \textbf{0.005} & \textbf{0.016} & 0.035 & \textbf{0.000} & \textbf{0.001} & 0.029 & \textbf{0.000} & \textbf{0.001} & 0.015 & 0.058 & 0.077 \\
 & Swin & / & / & / & 0.013 & 0.113 & 0.115 & 0.037 & \textbf{0.002} & \textbf{0.008} & 0.030 & \textbf{0.005} & \textbf{0.015} & 0.004 & 0.506 & 0.354 & 0.022 & \textbf{0.016} & \textbf{0.032} & 0.039 & \textbf{0.004} & \textbf{0.014} & 0.037 & \textbf{0.000} & \textbf{0.002} & 0.012 & 0.156 & 0.144 \\
 & VM & / & / & / & 0.012 & \textbf{0.049} & 0.070 & 0.035 & \textbf{0.000} & \textbf{0.001} & 0.035 & \textbf{0.000} & \textbf{0.002} & 0.011 & 0.086 & 0.103 & 0.027 & \textbf{0.001} & \textbf{0.003} & 0.048 & \textbf{0.000} & \textbf{0.000} & 0.039 & \textbf{0.000} & \textbf{0.000} & 0.020 & \textbf{0.019} & \textbf{0.036} \\

\hline
\multirow{3}{*}{OPL} & GT & / & / & / & 0.001 & 0.751 & 0.475 & 0.004 & 0.350 & 0.276 & 0.004 & 0.152 & 0.143 & 0.001 & 0.817 & 0.501 & 0.030 & \textbf{0.006} & \textbf{0.016} & 0.017 & 0.072 & 0.089 & 0.014 & \textbf{0.019} & \textbf{0.036} & 0.015 & 0.060 & 0.078 \\
 & Swin & / & / & / & 0.015 & \textbf{0.033} & 0.052 & 0.033 & \textbf{0.000} & \textbf{0.000} & 0.030 & \textbf{0.000} & \textbf{0.000} & 0.009 & 0.096 & 0.106 & 0.028 & \textbf{0.004} & \textbf{0.014} & 0.069 & \textbf{0.000} & \textbf{0.000} & 0.037 & \textbf{0.000} & \textbf{0.001} & 0.015 & 0.068 & 0.085 \\
 & VM & / & / & / & 0.002 & 0.516 & 0.359 & 0.015 & \textbf{0.002} & \textbf{0.010} & 0.011 & \textbf{0.008} & \textbf{0.020} & 0.001 & 0.720 & 0.459 & 0.026 & \textbf{0.001} & \textbf{0.006} & 0.069 & \textbf{0.000} & \textbf{0.000} & 0.028 & \textbf{0.000} & \textbf{0.002} & 0.014 & 0.054 & 0.072 \\

\hline
\multirow{3}{*}{ONL+IS} & GT & / & / & / & 0.001 & 0.624 & 0.408 & -0.003 & 0.271 & 0.226 & -0.004 & 0.091 & 0.106 & -0.006 & \textbf{0.004} & \textbf{0.014} & 0.004 & \textbf{0.011} & \textbf{0.022} & 0.005 & 0.088 & 0.104 & 0.004 & 0.205 & 0.182 & 0.001 & 0.818 & 0.501 \\
 & Swin & / & / & / & 0.001 & 0.556 & 0.381 & -0.005 & 0.125 & 0.123 & -0.004 & 0.115 & 0.116 & -0.006 & \textbf{0.037} & 0.056 & 0.006 & \textbf{0.005} & \textbf{0.015} & 0.005 & 0.129 & 0.127 & 0.004 & 0.132 & 0.128 & 0.001 & 0.560 & 0.381 \\
 & VM & / & / & / & 0.002 & 0.337 & 0.270 & -0.003 & 0.291 & 0.239 & -0.003 & 0.203 & 0.181 & -0.004 & 0.139 & 0.133 & 0.006 & \textbf{0.003} & \textbf{0.014} & 0.005 & 0.111 & 0.115 & 0.005 & 0.064 & 0.083 & 0.002 & 0.355 & 0.277 \\

\hline
\multirow{3}{*}{EZ} & GT & 0.003 & 0.089 & 0.104 & 0.027 & 0.073 & 0.089 & 0.019 & \textbf{0.004} & \textbf{0.015} & 0.018 & 0.073 & 0.089 & 0.007 & 0.105 & 0.110 & 0.066 & \textbf{0.006} & \textbf{0.017} & 0.062 & \textbf{0.000} & \textbf{0.003} & 0.060 & \textbf{0.020} & \textbf{0.036} & 0.062 & \textbf{0.009} & \textbf{0.020} \\
 & Swin & 0.004 & 0.165 & 0.151 & 0.010 & 0.264 & 0.222 & 0.034 & \textbf{0.009} & \textbf{0.020} & 0.016 & 0.105 & 0.110 & 0.007 & 0.181 & 0.163 & 0.060 & \textbf{0.010} & \textbf{0.022} & 0.076 & \textbf{0.009} & \textbf{0.020} & 0.089 & \textbf{0.011} & \textbf{0.022} & 0.059 & 0.056 & 0.075 \\
 & VM & 0.005 & \textbf{0.038} & 0.057 & 0.033 & 0.082 & 0.099 & 0.032 & \textbf{0.004} & \textbf{0.014} & 0.021 & \textbf{0.038} & 0.057 & 0.008 & 0.110 & 0.115 & 0.101 & \textbf{0.000} & \textbf{0.002} & 0.094 & \textbf{0.000} & \textbf{0.002} & 0.100 & \textbf{0.003} & \textbf{0.014} & 0.044 & \textbf{0.039} & 0.059 \\

\hline
\multirow{3}{*}{OS} & GT & 0.000 & 0.990 & 0.594 & -0.018 & 0.249 & 0.213 & -0.019 & 0.225 & 0.195 & -0.010 & 0.556 & 0.381 & -0.025 & 0.092 & 0.106 & -0.011 & 0.590 & 0.391 & -0.023 & 0.317 & 0.258 & -0.008 & 0.709 & 0.454 & -0.006 & 0.782 & 0.491 \\
 & Swin & -0.036 & \textbf{0.006} & \textbf{0.016} & -0.044 & \textbf{0.031} & 0.051 & -0.048 & \textbf{0.028} & \textbf{0.047} & -0.041 & 0.076 & 0.092 & -0.043 & \textbf{0.032} & 0.051 & -0.015 & 0.617 & 0.405 & -0.041 & 0.151 & 0.143 & -0.035 & 0.224 & 0.195 & -0.025 & 0.261 & 0.220 \\
 & VM & -0.031 & \textbf{0.007} & \textbf{0.017} & -0.015 & 0.476 & 0.340 & -0.019 & 0.353 & 0.277 & -0.019 & 0.388 & 0.291 & -0.031 & 0.133 & 0.128 & 0.034 & 0.252 & 0.214 & 0.017 & 0.580 & 0.387 & 0.009 & 0.763 & 0.480 & 0.014 & 0.649 & 0.419 \\

\hline
\multirow{3}{*}{RPE} & GT & -0.023 & \textbf{0.007} & \textbf{0.017} & 0.003 & 0.796 & 0.494 & -0.021 & \textbf{0.050} & 0.070 & -0.009 & 0.496 & 0.351 & -0.012 & 0.319 & 0.258 & 0.022 & 0.151 & 0.143 & -0.013 & 0.315 & 0.258 & 0.015 & 0.344 & 0.273 & 0.012 & 0.408 & 0.302 \\
 & Swin & -0.023 & \textbf{0.008} & \textbf{0.020} & -0.016 & 0.157 & 0.144 & -0.020 & 0.103 & 0.110 & -0.007 & 0.568 & 0.383 & -0.028 & 0.059 & 0.078 & 0.012 & 0.412 & 0.302 & -0.013 & 0.413 & 0.302 & 0.013 & 0.459 & 0.333 & 0.012 & 0.554 & 0.381 \\
 & VM & 0.000 & \textbf{0.006} & \textbf{0.017} & 0.001 & \textbf{0.006} & \textbf{0.016} & -0.003 & 0.849 & 0.516 & 0.004 & 0.797 & 0.494 & 0.000 & \textbf{0.006} & \textbf{0.016} & 0.031 & 0.067 & 0.085 & -0.004 & 0.338 & 0.270 & 0.015 & 0.469 & 0.338 & 0.000 & \textbf{0.006} & \textbf{0.016} \\

\hline
\multirow{3}{*}{Fluid} & GT & 0.000 & \textbf{0.020} & \textbf{0.036} & 0.000 & 0.155 & 0.144 & 0.000 & \textbf{0.010} & \textbf{0.021} & 0.000 & \textbf{0.001} & \textbf{0.003} & 0.000 & 0.115 & 0.116 & 0.000 & \textbf{0.023} & \textbf{0.040} & 0.000 & \textbf{0.043} & 0.062 & 0.000 & \textbf{0.024} & \textbf{0.041} & 0.000 & 0.066 & 0.084 \\
 & Swin & 0.000 & \textbf{0.044} & 0.064 & 0.000 & 0.388 & 0.291 & 0.000 & \textbf{0.004} & \textbf{0.015} & 0.000 & \textbf{0.000} & \textbf{0.002} & 0.000 & 0.272 & 0.226 & 0.000 & 0.112 & 0.115 & 0.000 & \textbf{0.027} & \textbf{0.044} & 0.000 & 0.180 & 0.163 & 0.000 & 0.282 & 0.234 \\
 & VM & 0.000 & \textbf{0.042} & 0.062 & 0.000 & 0.494 & 0.351 & 0.000 & \textbf{0.026} & \textbf{0.044} & 0.000 & \textbf{0.004} & \textbf{0.014} & 0.000 & 0.412 & 0.302 & 0.000 & \textbf{0.032} & 0.052 & 0.000 & 0.336 & 0.270 & 0.000 & 0.216 & 0.189 & 0.000 & 0.507 & 0.354 \\

\hline
\multirow{3}{*}{HRF} & GT & 0.000 & \textbf{0.001} & \textbf{0.004} & 0.000 & \textbf{0.004} & \textbf{0.015} & 0.000 & 0.053 & 0.072 & 0.000 & \textbf{0.001} & \textbf{0.003} & 0.000 & \textbf{0.041} & 0.060 & 0.000 & 0.093 & 0.106 & 0.000 & \textbf{0.005} & \textbf{0.015} & 0.000 & \textbf{0.024} & \textbf{0.041} & 0.000 & 0.052 & 0.072 \\
 & Swin & 0.000 & \textbf{0.000} & \textbf{0.002} & 0.000 & \textbf{0.018} & \textbf{0.035} & 0.000 & \textbf{0.001} & \textbf{0.006} & 0.000 & \textbf{0.000} & \textbf{0.001} & 0.000 & \textbf{0.000} & \textbf{0.002} & 0.000 & 0.381 & 0.289 & 0.000 & \textbf{0.000} & \textbf{0.003} & 0.000 & \textbf{0.004} & \textbf{0.014} & 0.000 & 0.095 & 0.106 \\
 & VM & 0.000 & 0.102 & 0.110 & 0.000 & 0.426 & 0.310 & 0.000 & \textbf{0.005} & \textbf{0.015} & 0.000 & \textbf{0.022} & \textbf{0.040} & 0.000 & \textbf{0.009} & \textbf{0.020} & 0.000 & 0.797 & 0.494 & 0.000 & \textbf{0.007} & \textbf{0.017} & 0.000 & 0.053 & 0.072 & 0.000 & 0.215 & 0.189 \\

\hline
\end{tabular}
}
\caption{Correlation between visual acuity and layer sector thickness for NPDR patients using a generalized linear model (GLM) while controlling for age, gender and diabetes duration. The thickness is generated from ground truth(GT), SwinUNETR(Swin), and VM-UNet(VM). The retinal region sector, effect size, and p-value before and after FDR correction are indicated. Significant p-values are highlighted in \textbf{bold}.} 
\label{tab:VA_NPDR}
\end{table}

\newpage
\begin{table}[htbp]
\centering
\scriptsize
\renewcommand{\arraystretch}{1.1}
\resizebox{\linewidth}{!}{
\begin{tabular}{|c|c|ccc|ccc|ccc|ccc|ccc|ccc|ccc|ccc|ccc|}
\hline
\multicolumn{2}{|c|}{} & \multicolumn{3}{c|}{C} & \multicolumn{3}{c|}{SI} & \multicolumn{3}{c|}{NI} & \multicolumn{3}{c|}{II} & \multicolumn{3}{c|}{TI} & \multicolumn{3}{c|}{SO} & \multicolumn{3}{c|}{NO} & \multicolumn{3}{c|}{IO} & \multicolumn{3}{c|}{TO} \\
\multicolumn{2}{|c|}{} & $\beta$ & p$_\mathrm{raw}$ & p$_\mathrm{FDR}$ & $\beta$ & p$_\mathrm{raw}$ & p$_\mathrm{FDR}$ & $\beta$ & p$_\mathrm{raw}$ & p$_\mathrm{FDR}$ & $\beta$ & p$_\mathrm{raw}$ & p$_\mathrm{FDR}$ & $\beta$ & p$_\mathrm{raw}$ & p$_\mathrm{FDR}$ & $\beta$ & p$_\mathrm{raw}$ & p$_\mathrm{FDR}$ & $\beta$ & p$_\mathrm{raw}$ & p$_\mathrm{FDR}$ & $\beta$ & p$_\mathrm{raw}$ & p$_\mathrm{FDR}$ & $\beta$ & p$_\mathrm{raw}$ & p$_\mathrm{FDR}$ \\
\hline
\multirow{3}{*}{RNFL} & GT & / & / & / & -0.004 & 0.526 & 0.962 & -0.004 & 0.584 & 0.991 & -0.006 & 0.436 & 0.929 & 0.001 & 0.922 & 0.993 & -0.004 & 0.590 & 0.991 & -0.008 & 0.104 & 0.647 & -0.005 & 0.297 & 0.838 & 0.022 & \textbf{0.015} & 0.478 \\
 & Swin & / & / & / & -0.004 & 0.653 & 0.993 & -0.001 & 0.872 & 0.993 & -0.006 & 0.500 & 0.959 & 0.005 & 0.769 & 0.993 & -0.003 & 0.673 & 0.993 & -0.002 & 0.659 & 0.993 & -0.003 & 0.636 & 0.993 & 0.025 & \textbf{0.004} & 0.478 \\
 & VM & / & / & / & -0.008 & 0.345 & 0.880 & -0.008 & 0.406 & 0.929 & -0.011 & 0.164 & 0.688 & 0.004 & 0.841 & 0.993 & -0.005 & 0.488 & 0.959 & -0.006 & 0.216 & 0.734 & -0.006 & 0.299 & 0.838 & 0.027 & \textbf{0.004} & 0.478 \\

\hline
\multirow{3}{*}{GCL+IPL} & GT & / & / & / & -0.013 & \textbf{0.015} & 0.478 & -0.007 & 0.209 & 0.723 & -0.007 & 0.242 & 0.762 & -0.008 & 0.140 & 0.688 & -0.009 & 0.336 & 0.873 & -0.008 & 0.343 & 0.880 & -0.016 & 0.057 & 0.616 & -0.011 & 0.099 & 0.647 \\
 & Swin & / & / & / & -0.012 & \textbf{0.027} & 0.616 & -0.003 & 0.524 & 0.962 & -0.007 & 0.249 & 0.764 & -0.008 & 0.129 & 0.688 & -0.006 & 0.571 & 0.991 & 0.000 & 0.961 & 0.994 & -0.010 & 0.205 & 0.723 & -0.013 & 0.063 & 0.616 \\
 & VM & / & / & / & -0.010 & \textbf{0.050} & 0.616 & -0.007 & 0.209 & 0.723 & -0.005 & 0.400 & 0.929 & -0.006 & 0.221 & 0.743 & -0.008 & 0.366 & 0.918 & -0.001 & 0.506 & 0.959 & -0.014 & 0.057 & 0.616 & -0.010 & 0.134 & 0.688 \\

\hline
\multirow{3}{*}{INL} & GT & / & / & / & 0.011 & 0.242 & 0.762 & 0.013 & 0.173 & 0.688 & 0.007 & 0.432 & 0.929 & 0.007 & 0.078 & 0.647 & 0.006 & 0.751 & 0.993 & 0.023 & 0.198 & 0.723 & -0.009 & 0.586 & 0.991 & 0.004 & 0.738 & 0.993 \\
 & Swin & / & / & / & 0.001 & 0.928 & 0.993 & 0.016 & 0.177 & 0.693 & 0.008 & 0.464 & 0.955 & 0.008 & \textbf{0.050} & 0.616 & -0.001 & 0.922 & 0.993 & 0.017 & 0.169 & 0.688 & -0.004 & 0.823 & 0.993 & 0.003 & 0.826 & 0.993 \\
 & VM & / & / & / & 0.000 & 0.973 & 0.994 & 0.013 & 0.179 & 0.693 & 0.010 & 0.229 & 0.757 & 0.006 & 0.163 & 0.688 & -0.002 & 0.902 & 0.993 & 0.021 & 0.237 & 0.762 & 0.003 & 0.851 & 0.993 & -0.001 & 0.923 & 0.993 \\

\hline
\multirow{3}{*}{OPL} & GT & / & / & / & -0.005 & 0.431 & 0.929 & 0.000 & 0.973 & 0.994 & 0.000 & 0.988 & 0.994 & -0.003 & 0.476 & 0.959 & -0.001 & 0.948 & 0.993 & -0.002 & 0.932 & 0.993 & 0.023 & 0.160 & 0.688 & -0.002 & 0.857 & 0.993 \\
 & Swin & / & / & / & 0.001 & 0.942 & 0.993 & 0.004 & 0.548 & 0.971 & 0.005 & 0.422 & 0.929 & -0.002 & 0.713 & 0.993 & -0.002 & 0.899 & 0.993 & 0.015 & 0.312 & 0.838 & 0.003 & 0.861 & 0.993 & 0.007 & 0.599 & 0.991 \\
 & VM & / & / & / & 0.002 & 0.660 & 0.993 & 0.002 & 0.670 & 0.993 & 0.003 & 0.435 & 0.929 & -0.000 & 0.866 & 0.993 & 0.003 & 0.843 & 0.993 & 0.020 & 0.303 & 0.838 & 0.012 & 0.543 & 0.971 & 0.003 & 0.768 & 0.993 \\

\hline
\multirow{3}{*}{ONL+IS} & GT & / & / & / & -0.002 & 0.400 & 0.929 & 0.001 & 0.873 & 0.993 & -0.006 & 0.150 & 0.688 & -0.006 & \textbf{0.036} & 0.616 & -0.001 & 0.736 & 0.993 & 0.004 & 0.249 & 0.764 & -0.001 & 0.887 & 0.993 & -0.002 & 0.500 & 0.959 \\
 & Swin & / & / & / & 0.000 & 0.954 & 0.993 & 0.003 & 0.448 & 0.929 & -0.002 & 0.593 & 0.991 & -0.006 & 0.101 & 0.647 & 0.001 & 0.850 & 0.993 & 0.000 & 0.900 & 0.993 & -0.001 & 0.890 & 0.993 & 0.000 & 0.059 & 0.616 \\
 & VM & / & / & / & -0.001 & 0.674 & 0.993 & 0.001 & 0.715 & 0.993 & -0.002 & 0.595 & 0.991 & -0.005 & 0.100 & 0.647 & 0.000 & 0.928 & 0.993 & 0.003 & 0.273 & 0.809 & 0.001 & 0.810 & 0.993 & -0.000 & 0.949 & 0.993 \\

\hline
\multirow{3}{*}{EZ} & GT & 0.001 & 0.757 & 0.993 & -0.014 & 0.547 & 0.971 & 0.003 & 0.705 & 0.993 & 0.002 & 0.891 & 0.993 & 0.003 & 0.606 & 0.991 & -0.011 & 0.770 & 0.993 & 0.006 & 0.835 & 0.993 & -0.021 & 0.528 & 0.962 & -0.009 & 0.689 & 0.993 \\
 & Swin & -0.000 & 0.950 & 0.993 & -0.003 & 0.912 & 0.993 & -0.001 & 0.975 & 0.994 & -0.001 & 0.965 & 0.994 & 0.011 & 0.693 & 0.993 & -0.043 & 0.312 & 0.838 & -0.026 & 0.515 & 0.959 & -0.079 & 0.100 & 0.647 & -0.002 & 0.804 & 0.993 \\
 & VM & -0.001 & 0.886 & 0.993 & -0.004 & 0.903 & 0.993 & 0.007 & 0.780 & 0.993 & -0.006 & 0.875 & 0.993 & 0.002 & 0.912 & 0.993 & -0.006 & 0.870 & 0.993 & -0.013 & 0.725 & 0.993 & -0.013 & 0.772 & 0.993 & -0.033 & 0.367 & 0.918 \\

\hline
\multirow{3}{*}{OS} & GT & -0.027 & 0.115 & 0.684 & -0.052 & 0.099 & 0.647 & -0.059 & 0.081 & 0.647 & -0.052 & 0.121 & 0.688 & -0.053 & 0.065 & 0.616 & -0.074 & 0.107 & 0.650 & -0.046 & 0.312 & 0.838 & -0.060 & 0.134 & 0.688 & -0.078 & 0.063 & 0.616 \\
 & Swin & -0.015 & 0.593 & 0.991 & -0.017 & 0.641 & 0.993 & -0.028 & 0.210 & 0.723 & -0.012 & 0.745 & 0.993 & -0.023 & 0.554 & 0.974 & -0.055 & 0.206 & 0.723 & -0.004 & 0.442 & 0.929 & -0.060 & 0.098 & 0.647 & -0.062 & 0.157 & 0.688 \\
 & VM & -0.015 & 0.626 & 0.993 & -0.030 & 0.405 & 0.929 & -0.036 & 0.276 & 0.809 & -0.025 & 0.477 & 0.959 & -0.029 & 0.443 & 0.929 & -0.061 & 0.138 & 0.688 & -0.015 & 0.606 & 0.991 & -0.043 & 0.320 & 0.849 & 0.016 & 0.144 & 0.688 \\

\hline
\multirow{3}{*}{RPE} & GT & -0.012 & 0.303 & 0.838 & -0.028 & 0.084 & 0.647 & -0.020 & 0.173 & 0.688 & -0.023 & 0.140 & 0.688 & -0.020 & 0.165 & 0.688 & -0.015 & 0.409 & 0.929 & -0.012 & 0.488 & 0.959 & -0.026 & 0.142 & 0.688 & -0.016 & 0.421 & 0.929 \\
 & Swin & -0.012 & 0.405 & 0.929 & 0.003 & 0.864 & 0.993 & 0.000 & 0.989 & 0.994 & 0.017 & 0.406 & 0.929 & -0.002 & 0.932 & 0.993 & -0.002 & 0.930 & 0.993 & 0.003 & 0.828 & 0.993 & -0.013 & 0.502 & 0.959 & -0.006 & 0.783 & 0.993 \\
 & VM & -0.008 & 0.651 & 0.993 & -0.015 & 0.509 & 0.959 & -0.001 & 0.937 & 0.993 & 0.009 & 0.671 & 0.993 & -0.008 & 0.744 & 0.993 & -0.000 & 0.992 & 0.994 & -0.006 & 0.705 & 0.993 & 0.003 & 0.922 & 0.993 & -0.025 & 0.335 & 0.873 \\

\hline
\multirow{3}{*}{Fluid} & GT & 0.000 & \textbf{0.013} & 0.478 & 0.000 & \textbf{0.006} & 0.478 & 0.000 & \textbf{0.031} & 0.616 & 0.000 & \textbf{0.023} & 0.616 & 0.000 & \textbf{0.039} & 0.616 & 0.000 & 0.515 & 0.959 & 0.000 & 0.191 & 0.723 & 0.000 & 0.170 & 0.688 & 0.000 & 0.170 & 0.688 \\
 & Swin & 0.000 & \textbf{0.013} & 0.478 & 0.000 & 0.088 & 0.647 & 0.000 & 0.059 & 0.616 & 0.000 & 0.104 & 0.647 & 0.000 & 0.064 & 0.616 & -0.000 & 0.203 & 0.723 & 0.000 & 0.240 & 0.762 & 0.000 & 0.675 & 0.993 & -0.000 & 0.951 & 0.993 \\
 & VM & 0.000 & \textbf{0.012} & 0.478 & 0.000 & \textbf{0.031} & 0.616 & 0.000 & \textbf{0.042} & 0.616 & 0.000 & \textbf{0.031} & 0.616 & 0.000 & \textbf{0.049} & 0.616 & 0.000 & 0.910 & 0.993 & 0.000 & 0.267 & 0.800 & 0.000 & 0.505 & 0.959 & 0.000 & 0.994 & 0.994 \\

\hline
\multirow{3}{*}{HRF} & GT & 0.000 & 0.697 & 0.993 & 0.000 & 0.447 & 0.929 & 0.000 & 0.153 & 0.688 & -0.000 & 0.633 & 0.993 & -0.000 & 0.731 & 0.993 & -0.000 & 0.261 & 0.792 & -0.000 & 0.747 & 0.993 & -0.000 & 0.987 & 0.994 & -0.000 & 0.847 & 0.993 \\
 & Swin & 0.000 & 0.566 & 0.988 & 0.000 & 0.688 & 0.993 & 0.000 & 0.080 & 0.647 & 0.000 & 0.489 & 0.959 & 0.000 & 0.427 & 0.929 & -0.000 & 0.882 & 0.993 & 0.000 & 0.390 & 0.929 & 0.000 & 0.418 & 0.929 & 0.000 & 0.616 & 0.993 \\
 & VM & 0.000 & 0.615 & 0.993 & 0.000 & 0.762 & 0.993 & 0.000 & 0.082 & 0.647 & 0.000 & 0.397 & 0.929 & 0.000 & 0.301 & 0.838 & -0.000 & 0.821 & 0.993 & 0.000 & 0.545 & 0.971 & 0.000 & 0.719 & 0.993 & 0.000 & 0.641 & 0.993 \\

\hline
\end{tabular}
}
\caption{Correlation between visual acuity and layer sector thickness for PDR patients using a generalized linear model (GLM) while controlling for age, gender and diabetes duration. The thickness is generated from ground truth(GT), SwinUNETR(Swin), and VM-UNet(VM). The retinal region sector, effect size, and p-value before and after FDR correction are indicated. Significant p-values are highlighted in \textbf{bold}.} 
\label{tab:VA_PDR}
\end{table}